\documentclass[structabstract]{aa}  
\usepackage{cancel,graphics,latexsym,epsfig,graphicx,amssymb,lscape,color,txfonts,natbib}     

\renewcommand{\d}[0]{{\rm d}}
\newcommand{\ave}[1]{\langle #1 \rangle}
\newcommand{\Ave}[1]{\big\langle #1\big\rangle}
\newcommand{\Ref}[1]{(\ref{#1})}

\newcommand{\mat}[1]{{\tens{#1}}}

\newcommand{\wtilde}[1]{\widetilde{#1}}

\newcommand{\ch}[1]{#1}

\begin{document}

\title{Improving three-dimensional mass mapping with weak
  gravitational lensing using galaxy clustering}

\author{Patrick Simon}

\institute{Argelander-Institut f\"ur Astronomie, Universit\"at Bonn, Auf dem
  H\"ugel 71, 53121 Bonn, Germany\\
  \email{psimon@astro.uni-bonn.de}}

\date{Received \today}

\authorrunning{Patrick Simon}
\titlerunning{Improved 3D lensing mass mapping}

\abstract {The weak gravitational lensing distortion of distant galaxy
  images (defined as sources) probes the projected large-scale matter
  distribution in the Universe. The availability of redshift
  information in galaxy surveys also allows us to recover the radial
  matter distribution to a certain degree.}
{To improve \ch{quality} in the mass mapping, we combine the lensing
  information with the spatial clustering of a population of galaxies
  that trace the matter density with a known galaxy bias \ch{(defined
    as tracers)}.}
{We construct a minimum-variance estimator for the 3D matter density
  that incorporates the angular distribution of galaxy tracers, which
  are coarsely binned in redshift. Merely all the second-order bias of
  the tracers has to be known, which can in principle be
  self-consistently constrained in the data by lensing
  techniques. \ch{This synergy introduces a new noise component
    because of the stochasticity in the matter-tracer density
    relation. We give a description of the stochasticity noise in the
    Gaussian regime, and we investigate the estimator characteristics
    analytically. We apply the estimator to a mock survey based on the
    Millennium Simulation.}}
{\ch{The estimator linearly mixes the individual lensing mass and
    tracer number density maps into a combined smoothed mass map. The
    weighting in the mix depends on the S/N of the individual maps and
    the correlation, $R$, between the matter and galaxy density. The
    weight of the tracers can be reduced by hand.}  For moderate
  mixing, the S/N in the mass map improves by a factor
  \mbox{$\sim2-3$} \ch{for $R\gtrsim0.4$. Importantly,} the systematic
  offset between a true and apparent mass peak distance (defined as
  $z$-shift bias) in a lensing-only map is eliminated, even for weak
  correlations of \mbox{$R\sim0.4$}.}
{\ch{If the second-order bias of tracer galaxies can be determined,
    the synergy technique potentially provides an option to improve
    redshift accuracy and completeness of the lensing 3D mass
    map. Herein, the aim is to visualise the spatial distribution of
    cluster-sized mass peaks. Our noise description of the estimator
    is accurate in the linear, Gaussian regime. However, its
    performance on sub-degree scales depends on the details in the
    galaxy bias mechanism and, hence, on the choice of the tracer
    population. Nonetheless, we expect that the mapping technique
    yields qualitatively reasonable results even for arcmin smoothing
    scales, as observed when this technique is applied to the mock
    survey with two different tracer populations.}}


\maketitle


\section{Introduction}
\label{firstpage}

The weak gravitational lensing effect is a well-established tool to
infer properties of the projected large-scale matter distribution
\citep[e.g.][]{2008PhR...462...67M,2006glsw.conf....1S,2006glsw.conf..269S}.
These therein exploited coherent shear distortions of distant galaxy
images (defined as sources) result from the continuous deflection of
light bundles by the intervening fluctuations in the large-scale
gravitational field, which are most prominent and detectable around
galaxy clusters. The lensing distortions probe the total matter
content in the Universe, which makes them an excellent tool for
studying the dark matter component, an essential ingredient of the
standard cosmological model of cold dark matter with a cosmological
constant \citep[$\Lambda$CDM, e.g.,][]{2003moco.book.....D}.

The shear distortion pattern can be translated into a map of projected
matter fluctuations. Early non-parametric mapping algorithms, which
were refined later to obtain optimised methods for finite fields,
achieved this only on the basis of a catalogue of source angular
positions and ellipticities
\citep[e.g.][]{1993ApJ...404..441K,2001A&A...374..740S}.  With the
advent of distance indicators of galaxies in wide field galaxy
surveys, the purely geometric relation between shear magnitude and
source (and lens) distance was incorporated into a new
three-dimensional (3D) lensing algorithm to also recover information
on the radial distribution of matter
\citep{hukeeton02,bacontay,2009MNRAS.399...48S,2011ApJ...727..118V,2011arXiv1111.6478L}. The
best studied methodologies so far utilise linear inversion techniques,
such as Wiener filtering or a radial matter-density eigenmode
decomposition with a suppression of low signal-to-noise (S/N)
modes. Owing to the relatively sparse and noisy sampling of the survey
area with background sources, however, the resulting maps are usually
very noisy, and significant detections are basically restricted to
mass peaks of a galaxy cluster scale that has only moderate redshift
accuracy. Moreover, the linear inversion utilises a radial smoothing
with a broad smoothing kernel that (a) smears out localised peaks in a
radial direction and (b) biases the peak distances (known as $z$-shift
bias; \citealt{2009MNRAS.399...48S}), which potentially renders the
resulting maps hard to interpret. To attain more realistic 3D maps,
the radial elongation of peaks inside the map can be mended by
regularising the inversion \citep{2011arXiv1111.6478L}, or by finding
the maximum likelihood positions of one or a few individual mass peaks
along the line-of-sight (l.o.s.) given the radial smoothing kernel and
radial density profile in the map
\citep{2012MNRAS.419..998S}. However, this does not alleviate the
principle problem of noisy maps and inaccurate peak distances. It
merely provides more realistic estimators for the 3D mass
map. Moreover, the noise properties of the maps are likely to be
complex in regularised, non-linear methods.

On the other hand, galaxy positions themselves are also tracers of the
3D matter density field and could therefore be employed to add extra
information to the matter density maps that are obtained from 3D
lensing. However, there are two complications here: (i) galaxies trace
the matter density field only up to a systematic mismatch, which is
generally dubbed \emph{galaxy bias}, and (ii) a sampling by galaxy
positions is affected by shot-noise
\citep[e.g.][]{1999ApJ...520...24D,2002sgd..book.....M}. The strategy
of this paper is to refine the minimum-variance estimator in
\citet{2009MNRAS.399...48S} (STH09 hereafter) for the 3D matter
density by adding the galaxy clustering information to the map making
process. Since the minimum-variance estimators
\citep{1995ApJ...449..446Z} require second-order statistics of the
input data to be specified, only the second-order bias parameters of
the galaxy tracers have to be known (Gaussian bias or linear
stochastic bias; \citealt{1999ApJ...520...24D}). The galaxy bias as a
function of scale and redshift could in principle be acquired in a
self-consistent approach from the data by using lensing techniques
\citep{1998ApJ...498...43S,1998A&A...334....1V,2003MNRAS.346..994,2003ApJ...594...33F,2012arXiv1202.6491J,
  2012arXiv1202.2046S}, or with lesser certainty from simulations
\citep{2001ApJ...558..520Y,2001MNRAS.320..289S,2004ApJ...601....1W}.
We therefore assume that it is basically known. The galaxy noise
covariance within the minimum-variance estimator takes care of the
galaxy sampling shot-noise.  \ch{The outline of this paper is as
  follows. The Sections \ref{sect:indep} and \ref{sect:recon1} present
  the details of the algorithm and a formalism to quantify its noise
  properties. We discuss the algorithm in the context of an idealised
  survey and then apply it to simulated data. In Section
  \ref{sect:mock}, we give details of the fiducial survey and the mock
  data. The results on the expected performance of the algorithm are
  presented in Section \ref{sect:results} and discussed in the final
  Section \ref{sect:discussion}.}

\section{Independent reconstructions}
\label{sect:indep}

\ch{We first consider the reconstruction of the matter density field
  and galaxy-number density field separately. The next section
  combines both into one 3D mass map.}

\subsection{Matter density on lens planes}
\label{sect:3dlensing}

We briefly summarise here the formalism already presented in STH09. We
adopt the exact notation that is employed therein.  For more details,
we refer the reader to this paper.

We split the source catalogue into \mbox{$i=1\ldots N_{\rm z}$}
sub-samples where a redshift probability distribution (p.d.f.) is
known. The complex ellipticities \citep{2001PhR...340..291B} of the
sources belonging to the $i$th sub-sample are binned on a 2D grid that
covers the field-of-view of the survey area. This ellipticity grid is
denoted by the vector $\vec{\epsilon}^{(i)}$, whose elements are the
sorted pixel values of the grid. Every sub-sample uses the same grid
geometry. The paper assumes that the weak lensing approximation is
accurate enough for the lensing catalogue on the whole. That is, for
the given source redshift and in the l.o.s. direction
$\vec{\theta}_i$, the complex ellipticity, $\vec{\epsilon}^{\rm s}$,
is an unbiased estimator of the shear distortion,
$\vec{\gamma}=\gamma_1+{\rm i}\gamma_2$,
\begin{equation}
  \ave{\vec{\epsilon}^{\rm s}}=
  \ave{\vec{\gamma}+\epsilon^{\rm i}}=
  \vec{\gamma}\;,
\end{equation}
where $\epsilon^{\rm i}$ denotes the intrinsic (unlensed) complex
ellipticity of a source image. Moreover, we assume a flat sky with a
Cartesian coordinate frame.

We slice the light-cone volume, where the matter distribution is
reconstructed, into $N_{\rm lp}$ slices. Within the slices we
approximate the matter density contrast as constant along the
line-of-sight. \ch{Every grid pixel defines a solid angle associated
  with a l.o.s. direction $\vec{\theta}$.} Thus, the fluctuations of
the matter density field inside a slice are fully described by the
angular distribution of mean density contrasts on a plane (lens plane)
and the width of the slice. The matter density contrast on a lens
plane, $\vec{\delta}_{\rm m}^{(i)}$, is binned with the same angular
grid as the source ellipticities. We represent the grids,
$\vec{\epsilon}^{(i)}$ and $\vec{\delta}_{\rm m}^{(i)}$, as vectors of
equally ordered pixel values. We refer to a particular pixel by
$\delta_{\rm m}^{(i)}(\vec{\theta}_j)$, where $\vec{\theta}_j$ is the
position of the pixel on the sky. Therefore, our algorithm represents
the 3D-matter density contrast as an approximation by a discrete set
of lens planes, which numerically limits the radial resolution, and a
discrete set of pixels on the sky, limiting the angular resolution.
The complete sets of ellipticity planes and lens planes are combined
inside vectors of grids:
\begin{eqnarray}  
  \vec{\epsilon}&=&\left[\vec{\epsilon}^{(1)},\ldots,\vec{\epsilon}^{(N_{\rm z})}\right]\;,\\
  \vec{\delta}_{\rm m}&=&\left[\vec{\delta}_{\rm
      m}^{(1)},\ldots,\vec{\delta}_{\rm m}^{(N_{\rm lp})}\right]\;,
\end{eqnarray}
respectively. The brackets, which group together the vector arguments,
should be understood as big vectors that are obtained by piling up all
embraced vectors on top of each other.

In the weak lensing regime, the (pixelised) lensing convergence
$\kappa^{(i)}(\vec{\theta}_j)$ in the lowest-order Born approximation
is the weighed projection of the density contrast on the lens planes:
\begin{equation}
  \vec{\kappa}=
  \left[
  \sum_{i=1}^{N_{\rm lp}}Q_{1i}\vec{\delta}_{\rm m}^{(i)},\ldots,
  \sum_{i=1}^{N_{\rm lp}}Q_{N_{\rm z}i}\vec{\delta}_{\rm m}^{(i)}
  \right]=:
  \mat{Q}\vec{\delta}_{\rm m}\;,
\end{equation}
where the coefficients $Q_{ij}$ express the response of the $i$th
convergence plane $\vec{\kappa}^{(i)}$ to the density contrast in the
$j$th lens plane. Namely,
\begin{equation}
  Q_{ij}=\frac{3\Omega_{\rm m}}{2D_{\rm H}^2}
  \int_{\chi_j}^{\chi_{j+1}}\d\chi\frac{\overline{W}^{(i)}(\chi)
    f_{\rm K}(\chi)}{a(\chi)}\;,
\end{equation}
where
\begin{equation}
  \overline{W}(\chi)=
  \int_\chi^\infty\d\chi^\prime\frac{f_{\rm
      K}(\chi^\prime-\chi)}{f_{\rm K}(\chi^\prime)}p^{(i)}_\chi(\chi^\prime)\;.
\end{equation}
The function $p^{(i)}_\chi(\chi)$ denotes the p.d.f. of sources in
comoving distance $\chi$ of the $i$th source sub-sample, and
\mbox{$[\chi_j,\chi_{j+1}[$} sets the comoving radial boundaries of
the $j$th matter slice.  We use $D_{\rm H}:=c/H_0$ for the Hubble
radius and $f_{\rm K}(\chi)$ for the (comoving) angular diameter
distance. The projection from a grid vector in $\vec{\delta}_{\rm
  m}$-space to a grid vector in $\vec{\kappa}$-space is hence denoted
by the operator $\mat{Q}$ that is acting on $\vec{\delta}_{\rm m}$.

The next step connects the convergence planes $\vec{\kappa}$ to the
shear planes by a convolution of the lensing convergence on the grid
\begin{equation}
  \vec{\gamma}=
  \left[
  \mat{P}_{\gamma\kappa}\vec{\kappa}^{(1)},\ldots,
  \mat{P}_{\gamma\kappa}\vec{\kappa}^{(N_{\rm z})}
  \right]=:
  \mat{P}_{\gamma\kappa}\vec{\kappa}\;,
\end{equation}
which introduces the operator $\mat{P}_{\gamma\kappa}$ to map
$\vec{\kappa}^{(i)}$ to the corresponding shear plane
$\vec{\gamma}^{(i)}$ \citep{hukeeton02}. In this sense,
$\mat{P}_{\gamma\kappa}$ performs a linear transformation from
$\vec{\kappa}$- to $\vec{\gamma}$-space.

Using this compact notation, we express the linear relation between
the matter density (contrast) on the lens planes and the observed,
binned ellipticity planes as:
\begin{equation}
  \label{eq:obsv}
  \vec{\epsilon}=
  \mat{P}_{\gamma\kappa}\mat{Q}\vec{\delta}_{\rm m}+\vec{n}_\gamma\;.
\end{equation}
Here, an additional vector $\vec{n}_\gamma$ denotes the binned
intrinsic ellipticties of the sources of all source sub-samples. In the
language of lensing, we consider this the noise term that dilutes the
shear signal $\mat{P}_{\gamma\kappa}\mat{Q}\vec{\delta}_{\rm m}$.

For the scope of this paper, possible correlations between shear and
intrinsic shapes are ignored \citep{2004PhRvD..70f3526H}. According to
STH09, minimum-variance estimator of $\vec{\delta}_{\rm m}$ in
Eq. \Ref{eq:obsv} is then
\begin{equation}
 \label{eq:dmestimator}
  \vec{\delta}_{\rm m,est}=
  \mat{S}_\delta\mat{Q}^{\rm t}\mat{P}^\dagger_{\gamma\kappa}
  \Big(\mat{N}^{-1}_\gamma\mat{P}_{\gamma\kappa}\mat{Q}\mat{S}_\delta\mat{Q}^{\rm t}\mat{P}^\dagger_{\gamma\kappa}+
    \alpha\mat{1}\Big)^{-1}\mat{N}_\gamma^{-1}\vec{\epsilon}\;.
\end{equation}
As the only input, the minimum-variance filter requires the signal
covariance $\mat{S}_\delta=\ave{\vec{\delta}_{\rm m}\vec{\delta}_{\rm
    m}^{\rm t}}$, which specifies the presumed two-point correlation
between pixel values of $\delta_{\rm m}^{(i)}(\vec{\theta})$ on the
lens plane(s) and the noise covariance
$\mat{N}_\gamma=\ave{\vec{n}_\gamma\vec{n}_\gamma^{\rm t}}$, which
quantifies the shear pixel noise variance and the correlation of noise
between different pixels. Pixels that contain no sources have infinite
noise. For the signal covariance, correlations between pixels that
belong to different lens planes are set to zero. We note here that the
signal covariance does not need to be the true signal covariance in
the data, although the reconstruction may be sub-optimal as to map
noise when it is not.

The signal covariance determines the degree of smoothing in the 3D
map. The smoothing is uniquely defined by the linear transformation
\begin{equation}
  \mat{B}_\delta:=
  \left(\alpha\mat{1}+\mat{S}_\delta\mat{N}_\delta^{-1}\right)^{-1}
  \mat{S}_\delta\mat{N}_\delta^{-1}~;~
  \mat{N}_\delta^{-1}:=\mat{Q}^{\rm t}\mat{P}^\dagger_{\gamma\kappa}
  \mat{N}_\gamma^{-1}\mat{P}_{\gamma\kappa}\mat{Q}\;,
\end{equation}
and can be utilised for a comparison of the map $\vec{\delta}_{\rm
  m,est}$ to a theoretical matter distribution $\vec{\delta}_{\rm
  m,th}$ by $\mat{B}_\delta\vec{\delta}_{\rm m,th}$
\citep{2012MNRAS.419..998S}. The radial smoothing is characterised by
a radial point-spread function (p.s.f.) of the filter (STH09). After
smoothing with the radial p.s.f., a peak in the true matter
distribution $\vec{\delta}_{\rm m,th}$ does not necessarily peak at
the same distance on average as in the smoothed map, which gives rise
to the so-called redshift bias or $z$-bias. Inside the filter, the
constant \mbox{$\alpha\in[0,1]$} tunes the level of smoothing by
rescaling the noise covariance.

From a practical point of view, the Wiener filter consists of a series
of linear operators that is applied step-by-step from the right to the
left on the grids (Appendix B of STH09). Within this process, the
signal covariance, $\mat{S}_\delta$, is a convolution or,
equivalently, a multiplication in Fourier space of Fourier modes,
$\tilde{f}(\vec{\ell})$, of the $i$th lens plane with the angular
signal power spectrum, $P_\delta^{(i)}(\ell)$, which is implicitly
defined by
\begin{equation}
  \Ave{\tilde{\delta}_{\rm m}^{(i)}(\vec{\ell})\tilde{\delta}_{\rm m}^{(i)}(\vec{\ell}^\prime)}
  =(2\pi)^2\delta_{\rm D}(\vec{\ell}+\vec{\ell}^\prime)P_\delta^{(i)}(|\vec{\ell}|)\;.
\end{equation}
We approximate the power spectrum by using Limber's equation in
Fourier space:
\begin{equation}
  \label{eq:pdelta}
  P_\delta^{(i)}(\ell)=
  \frac{|\tilde{F}(\ell)|^2}{(\Delta \chi_i)^2} \int_{\chi_i}^{\chi_{i+1}}
  \!\!\!\frac{\d\chi}{[f_{\rm K}(\chi)]^2}P_{\rm 3d}\left(\frac{\ell}{f_{\rm
        K}(\chi)},\chi\right)\;,
\end{equation}
where \mbox{$\Delta\chi_i:=\chi_{i+1}-\chi_i$}, $\tilde{F}(\ell)$ is
the Fourier transform of the pixel window function, $P_{\rm
  3d}(k,\chi)$ is the 3D matter-density power spectrum at radial
distance $\chi$ for wave-number $k$, and $\delta_{\rm D}(\vec{x})$ is
Dirac's delta function \citep{1992ApJ...388..272K}. We denote the
Fourier transforms of flat fields, $f(\vec{\theta})$, on the sky by
$\tilde{f}(\vec{\ell})$, which is defined by
\begin{equation}
  f(\vec{\theta})=
  \int\frac{\d^2\ell}{(2\pi)^2}\,\tilde{f}(\vec{\ell})
      {\rm e}^{+{\rm i}\vec{\theta}\cdot\vec{\ell}}~;~
  \tilde{f}(\vec{\ell})=
  \int\d^2\theta\,f(\vec{\theta})
  {\rm e}^{-{\rm i}\vec{\theta}\cdot\vec{\ell}}\;.
\end{equation}

\subsection{Galaxy numbers densities on lens planes}

To improve the information in the 3D matter map and to possibly
alleviate the $z$-shift bias, we add the information gained from
galaxy positions, which also probe the matter distribution (defined as
tracers).

In this section, however, we first visit the problem of mapping the
spatial galaxy number densities. For this purpose, we estimate the
number density of galaxies projected onto the previously defined lens
planes. Hence, we slice the full \emph{true} 3D galaxy distribution
into $N_{\rm lp}$ distance slices with distance limits
$[\chi_i,\chi_{i+1}[$. The galaxies are counted within each slice and
angular grid pixel of the solid angle $A_\omega$. Thereby, we receive
the galaxy number density \mbox{$n^{(i)}_{\rm
    g}(\vec{\theta}_j)=N^{(i)}(\vec{\theta}_j)/A_\omega$} in the
l.o.s. direction $\vec{\theta}_j$ of the $i$th slice, where
$N^{(i)}(\vec{\theta}_j)$ is the number of counted galaxies. We
compile the galaxy-number density values inside a grid vector
$\vec{n}^{(i)}_{\rm g}$, and we then arrange all grids inside a vector
of grids:
\begin{equation}
  \vec{n}_{\rm g}=
  \left[\vec{n}_{\rm g}^{(1)},\ldots,\vec{n}_{\rm g}^{(N_{\rm lp})}\right]\;.
\end{equation}

This number density distribution of galaxies is what the following
scheme seeks to recover from a galaxy sample with inaccurate distance
information. Towards this goal, we split the \emph{observed} galaxy
sample utilising their redshift estimators, $z_{\rm
  est}\in[z(\chi_i),z(\chi_{i+1})[$, into $N_{\rm lp}$ sub-samples
with known radial p.d.f. $p^{(i)}_{\rm f}(\chi)$; $z(\chi)$ denotes
the redshift corresponding to $\chi$. By projecting the $i$th sample
onto a 2D grid on the sky, one obtains the observed number density
distribution
\begin{equation}
  \label{eq:gdef}
  \eta^{(i)}_{\rm g}(\vec{\theta}_k)=
  \sum_{j=1}^{N_{\rm lp}}
  f_{\rm mask}(\vec{\theta}_k)p_{ij}\,
  n^{(j)}_{\rm g}(\vec{\theta}_k)
  =:
  \sum_{j=1}^{N_{\rm lp}}
  G_{ij}(\vec{\theta}_k)n_{\rm g}^{(j)}(\vec{\theta}_k)\;,
\end{equation}
where \mbox{$f_{\rm mask}\in\{0,1\}$} flags mask pixels ($=0$ for
mask), and
\begin{equation}
  p_{ij}:=
  \int_{\chi_j}^{\chi_{j+1}}\d\chi
  p^{(i)}_{\rm f}(\chi)
\end{equation}
is the probability \ch{that a galaxy inside $\eta_{\rm g}^{(i)}$
  belongs to the slice $j$.} Owing to the redshift errors and masking,
the observed distribution on the lens planes, $\eta^{(i)}_{\rm
  g}(\vec{\theta}_k)$, does not exactly match the true distribution
$n_{\rm g}^{(i)}(\vec{\theta}_k)$. Therefore, \mbox{$0\le
  G_{ij}(\vec{\theta}_k)\le 1$} denotes the expected fraction of
galaxies on the $j$th lens plane that is mapped onto the grid
$\eta_{\rm g}^{(i)}$. Because of masking, the total number of galaxies
is not necessarily conserved; that is \mbox{$\sum_{i=1}^{N_{\rm
      lp}}G_{ij}(\vec{\theta}_k)\ne1$}. By a proper arrangement of the
elements $G_{ij}(\vec{\theta}_k)$ inside a matrix $\mat{G}$, the
effect of $G_{ij}(\vec{\theta}_k)$ on the entire 3D grid $\vec{n}_{\rm
  g}$ can be written as
\begin{equation}
  \label{eq:obsv2}
  \vec{\eta}_{\rm g}= 
  \mat{G}\vec{n}_{\rm
    g}+\vec{\phi}_{\rm g}\;,
\end{equation}
where
\begin{equation}
  \vec{\eta}_{\rm g}=
  \left[\vec{\eta}_{\rm g}^{(1)},\ldots,\vec{\eta}_{\rm g}^{(N_{\rm lp})}\right]\;.
\end{equation}
We presume that galaxies sample an underlying smooth galaxy number
density by a discrete Poisson process
\citep[e.g.,][]{2002sgd..book.....M}. Therefore, the observable galaxy
counts sample the underlying galaxy number density $\vec{n}_{\rm g}$
up to shot-noise, which is here formally expressed by the noise
component $\vec{\phi}_{\rm g}$.

By analogy with the matter density $\vec{\delta}_{\rm m}$, we can find
an minimum-variance filter to estimate the true distribution of
galaxies on the lens planes; namely
\begin{equation}
  \label{eq:glxestimator}
  \vec{n}_{\rm g,est}=
  \mat{S}_{\rm g}\mat{G}^{\rm t}
  \Big(\mat{N}^{-1}_{\rm g}\mat{G}\mat{S}_{\rm g}\mat{G}^{\rm t}+
    \beta\mat{1}\Big)^{-1}
  \mat{N}^{-1}_{\rm g}\vec{\eta}_{\rm g}\;.
\end{equation}
As before, \mbox{$\mat{S}_{\rm g}=\ave{\vec{n}_{\rm g}\vec{n}_{\rm
      g}^{\rm t}}$} is the signal covariance, which is the angular
clustering two-point correlation function of the galaxies on the lens
planes, and \mbox{$\mat{N}_{\rm g}=\ave{\vec{\phi}_{\rm
      g}\vec{\phi}_{\rm g}^{\rm t}}$} denotes the shot-noise
covariance. The degree of smoothing by the Wiener filter is tunable by
using $\beta\in[0,1]$, which does not need to equal the parameter
$\alpha$ in Eq. \Ref{eq:dmestimator}. For the Poisson shot-noise
covariance, we adopt a diagonal noise covariance, $[\mat{N}_{\rm
  g}]_{ij}=0$ for \mbox{$i\ne j$}, with \mbox{$[\mat{N}_{\rm
    g}]_{ii}=\bar{\eta}_{\rm g}^{(k)}(\vec{\theta}_l)$} for unmasked
grid pixels $\vec{\theta}_l$, and infinite noise otherwise. The Wiener
filter in the given form requires the inverse noise covariance, such
that elements with infinite noise on the diagonal are zero. By
$\bar{\eta}_{\rm g}^{(k)}(\vec{\theta}_l)$, we denote the estimated
mean number density of galaxies in pixel $\vec{\theta}_l$ of the $k$th
sub-sample (see next section).

As for the matter density Wiener filter, a practical implementation of
the Wiener filter in Eq. \Ref{eq:glxestimator} consists of a series of
linear operations applied to $\vec{\eta}_{\rm g}$. The effect of
$\mat{S}_{\rm g}$ is to multiply every angular mode
$\tilde{f}(\vec{\ell})$ of the $i$th lens plane with the prior galaxy
power spectrum $P_{\rm g}^{(i)}(\ell)$, which we define relative to
the matter power spectrum using the galaxy bias factor
\mbox{$b^{(i)}(\ell)\ge0$} \citep[e.g.,][]{1998ApJ...500L..79T}:
\begin{equation}
  \label{eq:galaxybias}
  \Ave{\tilde{n}^{(i)}_{\rm g}(\vec{\ell})\tilde{n}^{(i)}_{\rm g}(\vec{\ell}^\prime)}=
  (2\pi)^2\delta_{\rm D}(\vec{\ell}+\vec{\ell}^\prime)\,
  \underbrace{\left[\bar{n}_{\rm g}^{(i)}b^{(i)}(|\vec{\ell}|)\right]^2
  P_\delta^{(i)}(|\vec{\ell}|)}_{P_{\rm g}^{(i)}(\ell)}\;,
\end{equation}
where $\bar{n}_{\rm g}^{(i)}$ denotes the true mean number density of
galaxies on the $i$th lens plane.  For this definition of the bias
factor, the shot-noise contribution to the galaxy power spectrum is
excluded as it is already accounted for in $\vec{\phi}_{\rm g}$.

The angular bias factor $b^{(i)}(\ell)$ is related to the 3D bias
factor $b(k,\chi)$, where $k$ is the comoving 3D wave-number, by a
projection that is approximated by Limber's equation:
\begin{eqnarray}
  \lefteqn{[b^{(i)}(\ell)]^2P_\delta^{(i)}(\ell)}\\
  &&=\nonumber
  \frac{|\tilde{F}(\ell)|^2}{[\Delta\chi_i]^2}
  \int_{\chi_i}^{\chi_{i+1}}\frac{\d\chi}{[f_{\rm K}(\chi)]^2}
  \,b^2\left(\frac{\ell}{f_{\rm K}(\chi)},\chi\right)
  P_{\rm 3d}\left(\frac{\ell}{f_{\rm K}(\chi)},\chi\right)\;;
\end{eqnarray}
$P_\delta^{(i)}(\ell)$ is given by Eq. \Ref{eq:pdelta}. For this
approximation, we assume that the number density of galaxies stays
constant as function of $\chi$ inside the slice.

\subsection{True mean galaxy numbers}

The true galaxy number densities $\bar{n}^{(i)}_{\rm g}$ in
Eq. \Ref{eq:galaxybias} have to be derived from the data itself. For
an estimator of $\bar{n}_{\rm g}^{(i)}$, we go back to
Eq. \Ref{eq:obsv2}, which relates the observed number of galaxies,
$\vec{\eta}_{\rm g}$, to the true number on the lens planes,
$\vec{n}_{\rm g}$. For an ensemble average of this relation, we expect
\begin{equation}
\bar{\vec{\eta}}_{\rm g}:=
\Ave{\vec{\eta}_{\rm g}}=
\mat{G}\bar{\vec{n}}_{\rm g}\;,
\end{equation}
wherein all elements $\bar{n}_{\rm g}^{(i)}(\vec{\theta}_j)$ equal the
same number $\bar{n}^{(i)}_{\rm g}$ owing to the statistical
homogeneity of the galaxy-number density fields, hence
\begin{equation}
  \label{eq:eta}
  \bar{\eta}_{\rm g}^{(i)}(\vec{\theta}_k)=
  \sum_{j=1}^{N_{\rm lp}}\mat{G}_{ij}(\vec{\theta}_k)
  \,\bar{n}_{\rm g}^{(j)}\;.
\end{equation}
Summing over all pixels with $N_{\rm p}$ in total of the $i$th tracer
sample yields
\begin{equation}
  \overline{\cal X}_{\rm
    g}^{(i)}:=
  \frac{1}{N_{\rm p}}\sum_{k=1}^{N_{\rm p}}\bar{\eta}_{\rm
    g}^{(i)}(\vec{\theta}_k)=
  \sum_{j=1}^{N_{\rm lp}}\overline{\mat{G}}_{ij}\,\bar{n}_{\rm g}^{(j)}\;,
\end{equation}
where
\begin{equation}
  \overline{\mat{G}}_{ij}:=\frac{1}{N_{\rm p}}\sum_{k=1}^{N_{\rm
      p}}G_{ij}(\vec{\theta}_k)
\end{equation}
averages $\mat{G}_{ij}$ over the area of the grid. Inverting the
former equation, gives 
\begin{equation}
  \label{eq:nbarest1}
  \bar{n}_{\rm g}^{(i)}=
  \sum_{j=1}^{N_{\rm
      lp}}\left[\bar{\mat{G}}^{-1}\right]_{ij}\,
  \overline{\cal X}_{\rm
    g}^{(j)}~~;~~
  \overline{\cal X}_{\rm g}^{(i)}\approx
  \frac{1}{N_{\rm p}}\sum_{k=1}^{N_{\rm
      p}}\eta^{(i)}_{\rm g}(\vec{\theta}_k)\;.
\end{equation}
For an unbiased estimator of $\overline{\cal X}_{\rm g}^{(i)}$ on the
right hand side, we insert the observed galaxy number densities, which
is \mbox{$\bar{\eta}_{\rm g}^{(i)}(\vec{\theta}_k)=\eta_{\rm
    g}^{(i)}(\vec{\theta}_k)$}. The value of $\bar{\eta}^{(i)}_{\rm
  g}(\vec{\theta}_k)$, which is utilised for the noise covariance
$\mat{N}_{\rm g}$ in the foregoing section, is computed from
Eq. \Ref{eq:eta} and the estimated $\bar{n}_{\rm g}^{(i)}$.

In the simple case of negligible redshift errors, we find
$\mat{G}_{ij}(\vec{\theta}_k)=\delta_{ij}^{\rm K}f_{\rm
  mask}(\vec{\theta}_k)$, where $\delta^{\rm K}_{ij}$ denotes the
Kronecker symbol.  In this case, we consequently find
\begin{equation}
  [\bar{\mat{G}}^{-1}]_{ij}=\delta^{\rm
  K}_{ij}\frac{N_{\rm p}}{N}\;
\end{equation}
for the number $N$ of unmasked pixels.  Moreover, we find
\mbox{$\bar{n}^{(i)}_{\rm g}=N_{\rm p}N^{(i)}_{\rm g}/(N\Omega)$} for
a number $N^{(i)}_{\rm g}$ of galaxies within the $i$th sub-sample and
a survey area $\Omega$. Thus, the galaxy number density $N^{(i)}_{\rm
  g}/\Omega$ is scaled up by $N_{\rm p}/N$ to account for the mask.

However, the estimator in Eq. \Ref{eq:nbarest1} has one caveat, since
$\bar{\eta}_{\rm g}^{(i)}$ is basically a convolution of $\bar{n}_{\rm
  g}^{(i)}$ with the redshift error of galaxies. A deconvolution
through $\bar{\mat{G}}^{-1}$ possibly results in oscillating and
negative values for $\bar{n}_{\rm g}^{(i)}$. We therefore regularise
Eq.  \Ref{eq:nbarest1} by a constrained solution of $\bar{n}_{\rm
  g}^{(i)}$ that maximises the likelihood:
\begin{equation}
  \ln{\cal L}(\bar{\vec{n}}_{\rm g}|\bar{\vec{\eta}}_{\rm g})=
  \left(\overline{\mat{G}}\bar{\vec{n}}_{\rm g}-\bar{\vec{\eta}}_{\rm
      g}\right)^{\rm t}
  \mat{N}_\eta^{-1}
  \left(\overline{\mat{G}}\bar{\vec{n}}_{\rm g}-\bar{\vec{\eta}}_{\rm g}\right)
\end{equation}
under the condition that \mbox{$\bar{n}_{\rm g}^{(i)}\ge0$} for all
$i$. We determine this solution numerically. The additional covariance
$\mat{N}_\eta$ can be used to give different weights to the observed
$\bar{\eta}_{\rm g}^{(i)}$ values, such as by weighing the number of
galaxies in each galaxy sample in order to account for the galaxy
shot-noise. For equal weights, we simply set
\mbox{$\mat{N}_\eta=\mat{1}$}.

\section{Combined reconstruction}
\label{sect:recon1}

\ch{In this section, we combine the information on the 3D matter
  density in the lensing data and the galaxy distribution.}

\subsection{Minimum-variance estimator}

Up to now, we have considered the galaxy number density and matter
density fields separately. However, $\vec{\eta}_{\rm g}$ contains
information about $\vec{\delta}_{\rm m}$ and vice versa, as galaxies
trace the matter distribution to a certain degree.  On a statistical
level, this relation is reflected by a non-vanishing cross-correlation,
\begin{equation}
  \mat{S}_{\delta{\rm
      g}}=
  \ave{\vec{\delta}_{\rm m}\vec{n}_{\rm g}^{\rm t}}\;,
\end{equation}
for pairs of pixels on the same lens plane, which has not entered our
formalism thus far.  Slices are thought to be wide enough, such that
correlations between pixels belonging to different lens plane are
negligible.

We combine the $\vec{\delta}_{\rm m}$- and $\vec{n}_{\rm g}$-grids
inside one new vector,
\begin{equation}
  \label{eq:eq30}
  \vec{s}:=\left[\vec{\delta}_{\rm m},\vec{n}_{\rm g}\right]\;.
\end{equation}
Eqs. \Ref{eq:obsv} and \Ref{eq:obsv2} relate $\vec{s}$ to the observed
shear and the tracer number density grids,
\begin{equation}
  \vec{d}:=\left[\vec{\gamma},\vec{\eta}_{\rm g}\right]\;,
\end{equation}
according to
\begin{equation}
  \vec{d}=
  \Big[\mat{P}_{\gamma\kappa}\mat{Q}\vec{\delta}_{\rm m},\mat{G}\vec{n}_{\rm
    g}\Big]+\vec{n}
  =:\mat{R}\vec{s}+\vec{n}\;,
\end{equation}
where the combined noise vector is
\begin{equation}
  \vec{n}:=\left[\vec{n}_\gamma,\vec{\phi}_{\rm g}\right]\;.
\end{equation}
In this compact notation, the action of a matrix
\begin{equation}
  \mat{A}=
  \left(\begin{array}{cc}
    \mat{A}_{11} & \mat{A}_{12}\\
    \mat{A}_{21} & \mat{A}_{22}
  \end{array}\right)\;,
\end{equation}
on a product vector $\vec{v}=[\vec{v}_1,\vec{v}_2]$ is defined as
\begin{equation}
  \mat{A}\vec{v}:=
  \big[
  \mat{A}_{11}\vec{v}_1+\mat{A}_{12}\vec{v}_2,
  \mat{A}_{21}\vec{v}_1+\mat{A}_{22}\vec{v}_2
  \big]\;.
\end{equation}
In this sense, the projection matrix $\mat{R}$ is
\begin{equation}
  \mat{R}=
  \left(\begin{array}{cc}
    \mat{P}_{\gamma\kappa}\mat{Q} & \mat{0}\\
    \mat{0} & \mat{G}
  \end{array}\right)\;.
\end{equation}

Following the usual assumptions of a minimum-variance filter, the
optimal filter for estimating $\vec{s}$ from $\vec{d}$ in this
combined problem is
\begin{equation}
  \label{eq:combinedest}
  \vec{s}_{\rm est}=
  \mat{S}\mat{R}^\dagger
  \big(\mat{N}_{\alpha\beta}^{-1}\mat{R}\mat{S}\mat{R}^\dagger+\mat{1}\big)^{-1}
  \mat{N}_{\alpha\beta}^{-1}\,\vec{d}\;,
\end{equation}
which uses the short-hand notations,
\begin{equation}
  \label{eq:shorthands}
  \mat{S}=
  \left(
  \begin{array}{cc}
    \mat{S}_\delta & \mat{S}_{\delta\rm g}\\
    \mat{S}_{\delta\rm g}^{\rm t} & \mat{S}_{\rm g}
  \end{array}\right)~;~
  \mat{N}_{\alpha\beta}=
  \left(
  \begin{array}{cc}
    \alpha\mat{N}_\gamma & \mat{0} \\
    \mat{0} & \beta\mat{N}_{\rm g}
  \end{array}
  \right)~;~
  \mat{R}^\dagger=
  \left(\begin{array}{cc}
    \mat{Q}^{\rm t}\mat{P}_{\gamma\kappa}^\dagger & \mat{0}\\
    \mat{0} & \mat{G}^{\rm t}
  \end{array}\right)\;.
\end{equation}
The galaxy shot-noise $\vec{\phi}_{\rm g}$ and the intrinsic
ellipticities of the sources, which are comprised in $\vec{n}_\gamma$,
are assumed to be uncorrelated. By choosing different tuning
parameters \mbox{$\alpha\ne\beta$}, the impact of the Wiener smoothing
can be adjusted independently for the matter and galaxy map.

The novelty of the combined reconstruction is that tracer number and
matter density maps exchange information, if the cross-correlation
matrix $\mat{S}_{\delta\rm g}$ is non-vanishing.  In a practical
implementation of the filter \Ref{eq:combinedest}, we apply step by
step linear operations to the grids stored inside $\vec{d}$ as
before. As with the previous operators $\mat{S}_\delta$ and
$\mat{S}_{\rm g}$, the application of $\mat{S}_{\delta\rm g}$ amounts
to a multiplication of angular grid modes with the cross-correlation
power spectrum, $P^{(i)}_{\delta\rm g}(\ell)$, determined by
\begin{equation}
  \Ave{\tilde{n}^{(i)}_{\rm g}(\vec{\ell})\tilde{\delta}_{\rm m}^{(i)}(\vec{\ell}^\prime)}
  =(2\pi)^2\delta_{\rm D}(\vec{\ell}+\vec{\ell}^\prime)\,
  \underbrace{\bar{n}_{\rm g}^{(i)}r^{(i)}(|\vec{\ell}|)b^{(i)}(|\vec{\ell}|)
    P_\delta^{(i)}(|\vec{\ell}|)}_{=P^{(i)}_{\delta\rm g}(\ell)}\;.
\end{equation}
\ch{(See the next section for details on the implementation.)} We
define $P^{(i)}_{\delta\rm g}(\ell)$ with respect to the matter power
spectrum $P^{(i)}_\delta(\ell)$ by employing the galaxy-matter
cross-correlation factor $r^{(i)}(\ell)$ \citep{1998ApJ...500L..79T}.
The angular function $r^{(i)}(\ell)$ is approximately related to the
3D correlation factor $r(k,\chi)$ according to
\begin{eqnarray}
  \lefteqn{r^{(i)}(\ell)b^{(i)}(\ell)P_\delta^{(i)}(\ell)}\\
  &&=\nonumber
  \frac{|\tilde{F}(\ell)|^2}{[\Delta\chi_i]^2}
  \int_{\chi_i}^{\chi_{i+1}}
  \!\!\!\!\frac{\d\chi}{[f_{\rm K}(\chi)]^2}
  \,r\left(k_\ell,\chi\right)
  b\left(k_\ell,\chi\right)
  P_{\rm 3d}\left(k_\ell,\chi\right)\;,
\end{eqnarray}
where $k_\ell:=\ell/f_{\rm K}(\chi)$.

To understand the mode of operation of the minimum-variance filter in
Eq. \Ref{eq:combinedest}, it is instructive to recast it into the
mathematically equivalent form:
\begin{equation}
  \label{eq:combinedwiener}
  \vec{s}_{\rm est}=
  \underbrace{\left(\mat{1}+\mat{S}\mat{N}^{-1}_{\delta n_{\rm g}}\right)^{-1}
  \mat{S}\mat{N}^{-1}_{\delta n_{\rm g}}}_{\rm Step-2}\times
  \underbrace{\mat{N}_{\delta n_{\rm
        g}}\mat{R}^\dagger\mat{N}_{\alpha\beta}^{-1}}_{\rm Step-1}\,\vec{d}\;,
\end{equation}
where $\mat{N}^{-1}_{\delta n_{\rm
    g}}:=\mat{R}^\dagger\mat{N}_{\alpha\beta}^{-1}\mat{R}$.  Step-1
involves no Wiener smoothing to construct the maps; no matrix
$\mat{S}$ is involved in this step. As this is usually too noisy, we
apply an additional smoothing to these maps by virtue of the Wiener
filter in Step-2. This filter linearly combines and averages pixel
values in the maps based upon the expected S/N in the unbiased
maps. It is Step-2, the analogue of the matrix $\mat{B}_\delta$ in
Sect. \ref{sect:3dlensing}, that introduces biases into the maps,
especially through a radial smoothing. Moreover, only Step-2 formally
mixes pixels from the mass map and the tracer number density map by
means of the off-diagonal matrix $\mat{S}_{\delta\rm g}$. Therefore,
Step-1 makes independent mass and tracer maps that are only later
combined in Step-2, according to our prior knowledge of their
correlation.  Setting \mbox{$\alpha=\beta=0$} results in a unity
matrix for Step-2 or no smoothing.

\ch{Analogous to a lensing-only reconstruction,} the Wiener filter
thus applies a radial and transverse smoothing to the map to increase
the signal-to-noise ratio.  The smoothing makes the maps biased
estimators of the matter and galaxy-number density fields.  The
smoothing is, however, uniquely defined by
\begin{equation}
  \label{eq:filter}
  \mat{B}:=\big(\mat{1}+\mat{S}\mat{N}_{\delta n_{\rm
      g}}^{-1}\big)^{-1}\mat{S}\mat{N}_{\delta n_{\rm g}}^{-1}\;,
\end{equation}
which and can be applied to theoretical maps of the matter and galaxy
number density for a quantitative comparison to the data.

\subsection{Fourier space representation}
\label{sect:fourierrep}

For shear and galaxy number noise homogeneous over infinite grids with
no gaps, the estimator in Eq.  \Ref{eq:combinedest} takes a simple
form in Fourier space. Under these idealistic conditions, the angular
modes of all lens planes combine to
\begin{equation}
  \tilde{\vec{s}}(\vec{\ell})=
  \left[\vec{\tilde{\delta}}_{\rm m}(\vec{\ell}),\vec{\tilde{n}}_{\rm
      g}(\vec{\ell})\right]\;,
\end{equation}
which are only linear functions of the $\eta_{\rm g}$- and
$\gamma$-modes of the same $\vec{\ell}$; there is no mixing between
modes of different $\ell$. Therefore, a reconstruction is then done
most easily in Fourier space by
\begin{equation}
\label{eq:combinedest2}
  \wtilde{\vec{s}}_{\rm est}(\vec{\ell})= 
  \wtilde{\mat{S}}_{\ell}
  \wtilde{\mat{R}}_{\ell}^\dagger
  \big(
  \wtilde{\mat{R}}_{\ell}^{}
  \wtilde{\mat{S}}_{\ell}
  \wtilde{\mat{R}}_{\ell}^\dagger+
  \wtilde{\mat{N}}_{\alpha\beta}\big)^{-1}
  \wtilde{\vec{d}}(\vec{\ell})\;,
\end{equation}
where $\wtilde{\vec{d}}(\vec{\ell})=
\left[\wtilde{\vec{\gamma}}(\vec{\ell}),\wtilde{\vec{\eta}}_{\rm
    g}(\vec{\ell})\right]$ are the observable input grids.  The tuned
covariance matrix of the (homogeneous) noise is
\begin{equation}
  \wtilde{\mat{N}}_{\alpha\beta}={\rm diag}\left\{
  \frac{\alpha\sigma_\epsilon^2}{\bar{n}_{\rm s}^{(1)}},\ldots,
  \frac{\alpha\sigma_\epsilon^2}{\bar{n}_{\rm s}^{(N_z)}},
  \beta\bar{\eta}_{\rm g}^{(1)},\ldots,\beta\bar{\eta}_{\rm g}^{(N_{\rm lp})}\right\}\;,
\end{equation}
where $\bar{n}_{\rm s}^{(i)}$ is the mean \emph{source} number density
of the $i$th source sample (out of in total $N_z$);
\mbox{$\sigma^2_\epsilon=\ave{\epsilon^{\rm i}[\epsilon^{\rm
      i}]^\ast}$} is their intrinsic shape noise variance, and
$\bar{\eta}_{\rm g}^{(i)}$ is the Poisson shot-noise power (white
noise). Possible noise contributions owing to intrinsic alignments of
sources are ignored here, hence $\wtilde{\mat{N}}_{\alpha\beta}$ has
no off-diagonal elements. Furthermore, one has
\begin{equation}
  \wtilde{\mat{R}}_{\ell}=
  \left(\begin{array}{cc}
    D(\vec{\ell})\mat{Q} & \mat{0} \\
    \mat{0} & \overline{\mat{G}}
    \end{array}\right)~;~
  \wtilde{\mat{R}}_{\ell}^\dagger=
  \left(\begin{array}{cc}
    D^\ast(\vec{\ell})\mat{Q}^{\rm t} & \mat{0} \\
    \mat{0} & \overline{\mat{G}}^{\rm t}
    \end{array}\right)\;,
\end{equation}
where $D(\vec{\ell})=\vec{\ell}/\vec{\ell}^\ast$
\citep{1993ApJ...404..441K}. For \mbox{$\vec{\ell}=0$}, we set
\mbox{$D(\vec{\ell})=0$}. Here, $\mat{G}$ does not depend on
$\vec{\theta}_k$. The signal covariance is
\begin{equation}
  \label{eq:sigcov}
  \wtilde{\mat{S}}_{\ell}=
  \left(\begin{array}{cc}
    \wtilde{\mat{S}}_\delta(\ell) & \wtilde{\mat{S}}_{\delta\rm g}(\ell) \\
    \wtilde{\mat{S}}_{\delta\rm g}(\ell) & \wtilde{\mat{S}}_{\rm g}(\ell)
  \end{array}\right) 
\end{equation}
with
\begin{eqnarray}
  \wtilde{\mat{S}}_\delta(\ell)&=&
  {\rm diag}\left\{P_\delta^{(1)}(\ell),\ldots,
    P_\delta^{(i)}(\ell),\ldots,P_\delta^{(N_{\rm lp})}(\ell)\right\}\;,\\
  \wtilde{\mat{S}}_{\delta\rm g}(\ell)&=&
  {\rm diag}\left\{
    \ldots,\bar{n}^{(i)}_{\rm g}r^{(i)}(\ell)b^{(i)}(\ell)P_\delta^{(i)}(\ell),\ldots\right\}\;,\\
  \wtilde{\mat{S}}_{\rm g}(\ell)&=&
  {\rm diag}\left\{
    \ldots,
    [\bar{n}^{(i)}_{\rm g}b^{(i)}(\ell)]^2
    P_\delta^{(i)}(\ell),\ldots\right\}\;.
\end{eqnarray}

Because of the diagonal structure of the last three matrices, the
matrix $\wtilde{\mat{S}}_{\ell}$ acting on a vector
$\wtilde{\vec{v}}(\vec{\ell})$ actually only mixes the matter and
tracer density modes from the same lens plane and of the same wave
vector $\vec{\ell}$. Thus, rearranging the modes inside
$\wtilde{\vec{v}}(\vec{\ell})$ and pairing together matter and tracer
density modes of the same lens plane render
$\wtilde{\mat{S}}_{\vec{\ell}}$ a diagonal block matrix, such that
\begin{equation}
  \wtilde{\mat{S}}_{\ell}={\rm diag}\left\{
    \wtilde{\mat{S}}^{(1)}_{\ell},\dots,\wtilde{\mat{S}}^{(N_{\rm lp})}_{\ell}\right\}
\end{equation}
with $2\times2$-blocks
\begin{equation}
  \wtilde{\mat{S}}^{(i)}_{\ell}=\left(
    \begin{array}{cc}
      1 & \bar{n}_{\rm g}^{(i)}r^{(i)}(\ell)b^{(i)}(\ell) \\
      \bar{n}_{\rm g}^{(i)}r^{(i)}(\ell)b^{(i)}(\ell) & 
      [\bar{n}_{\rm g}^{(i)}b^{(i)}(\ell)]^2      
    \end{array}\right)\times P_\delta^{(i)}(\ell)
\end{equation}
on the diagonal. This structure is useful when implementing the action
of $\mat{S}$ in \Ref{eq:combinedest} in practise. Clearly, modes will
not affect each other when \mbox{$r^{(i)}(\ell)=0$} with no
improvement by the synergy of lensing and galaxy tracers.

\subsection{Radial point spread function}
\label{sect:rpsf}

The radial p.s.f. is the average sight-line profile of a single mass
peak in the smoothed matter map. Ideally, the p.s.f. spikes at the
true mass peak redshift (no $z$-shift bias).  \ch{In reality, however,
  a $z$-shift bias is one of the main nuisances in Wiener
  reconstructions with 3D lensing data.} We assume a homogeneous
survey, where the choice of the l.o.s. direction $\vec{\theta}$ is
irrelevant. We hence arbitrarily pick \mbox{$\vec{\theta}=0$} as a
reference direction and omit the pixel index $\vec{\theta}$ in the
following.

We consider a singular test peak with a profile of \mbox{$\delta_{\rm
    m}(\chi)=A_{\rm p}\delta_{\rm D}(\chi-\bar{\chi}_i)$} in the
un-smoothed map; \ch{$A_{\rm p}$ is the peak amplitude.} It is located
at the distance $\bar{\chi}_i:=(\chi_i+\chi_{i+1})/2$ of the $i$th
lens plane. For circular pixels with angular radius $\Theta_{\rm s}$,
the pixel value of this peak is in the un-smoothed map
\begin{equation}
  F(\vec{\Delta\theta})=
  \left\{\begin{array}{ll}
      A_{\rm p}(\pi\Theta_{\rm s}^2)^{-1} & {\rm for}~|\vec{\Delta\theta}|\le\Theta_{\rm
        s}\\
      0 & {\rm otherwise}
  \end{array}\right.\;,
\end{equation}
or in the Fourier space,
\begin{equation}
  \wtilde{F}(\ell\Theta_{\rm s})=
  \frac{2A_{\rm p}J_1(\ell\Theta_{\rm s})}{\ell\Theta_{\rm s}}\;.
\end{equation}
\ch{By $J_n(x)$, we denote the spherical Bessel functions of the first
  kind.} Because of the linearity of the reconstruction algorithm, the
peak amplitude is unimportant for the shape of the radial p.s.f. We
therefore simply set \mbox{$A_{\rm p}=1$}.  Unlike the discussion in
STH09 for calculating the radial p.s.f., we also have to factor in the
tracer number density on the $i$th lens plane here (and same direction
$\vec{\theta}$). As this is a random variable for \mbox{$r(\ell)\ne
  1$}, we define the p.s.f. as the radial density profile in the
smoothed map given a matter peak $\tilde{\delta}_{\rm
  m}^{(i)}(\ell)=\wtilde{F}(\Theta_{\rm s}\ell)$ on the $i$th lens
plane that is marginalised over the tracer density $\tilde{n}_{\rm
  g}^{(i)}(\ell)$. This is associated with the mass peak.  This
conditional mean tracer number density is given by
\begin{equation}
  \label{eq:meannell}
  \tilde{n}^{(i)}_{\rm g}(\ell;F):=
  \Ave{\tilde{n}_{\rm g}^{(i)}(\ell)\Big|
    \tilde{\delta}_{\rm m}^{(i)}(\ell)}_n\approx
  \bar{n}_{\rm g}^{(i)}\wtilde{F}(\Theta_{\rm s}\ell)r^{(i)}(\ell)b^{(i)}(\ell)\;,
\end{equation}
where the conditional ensemble average 
\begin{equation}
  \ave{x|y}_n=
  \frac{\int\d x\,P(x,y)x}{\int\d x\,P(x,y)}
\end{equation}
is taken over all realisations of the tracer density field and
$P(x,y)$ denotes the bivariate p.d.f. of the tracer number density $x$
and the matter density $y$. The expression on the r.h.s. in the
Eq. \Ref{eq:meannell} is exact only for Gaussian statistics, which is
assumed here as lowest-order approximation (\ch{Appendix
  \ref{ap:mockdatagauss}}). For differing statistics, such as a
log-normal tracer density field \citep{1991MNRAS.248....1C}, we have
to expect deviations from this expression.  Evidently, the conditional
average will vanish if the correlation factor is
\mbox{$r^{(i)}(\ell)=0$}. The average tracer number density about a
mass peak vanishes in this case.

According to this definition, the radial p.s.f. equals the average
sight-line density profile (analogous to Eq. 77 of STH09):
\begin{equation}
  \label{eq:radialpsf}
  \left[\bar{\vec{\delta}}_{\rm m}(\Theta_{\rm
      s}),\bar{\vec{n}}_{\rm g}(\Theta_{\rm s})\right]=
  \int_0^\infty\frac{\d\ell\ell}{2\pi}\,
  \wtilde{F}(\ell\Theta_{\rm s})
  \,\wtilde{\mat{W}}_{\ell}
  \left[\vec{\tilde{\delta}}_{\rm m}(\ell),\vec{\tilde{n}}_{\rm g}(\ell)\right]\;,
\end{equation}
where the Wiener filter $\wtilde{\mat{W}}_{\ell}$ is given in
Eq. \Ref{eq:fpfilter2} and the vectors $\vec{\tilde{\delta}}_{\rm
  m}(\ell)$ and $\vec{\tilde{n}}_{{\rm g}}(\ell)$ ($N_{\rm lp}$
elements) vanish everywhere except in their $i$th element that equals
$1$ and $\bar{n}_{\rm g}^{(i)}r^{(i)}(\ell)b^{(i)}(\ell)$,
respectively. The elements of the vector $\bar{\vec{\delta}}_{\rm
  m}(\Theta_{\rm s})$ encapsulate the radial p.s.f. of the matter map,
and the radial p.s.f. of the tracer number density map in the case of
$\bar{\vec{n}}_{\rm g}(\Theta_{\rm s})$. The former is the focus in
the following.

\subsection{Map signal-to-noise}
\label{sect:mapsn}

With the estimator \Ref{eq:combinedest2} at hand, we forecast the S/N
of the matter and tracer number density modes as a function of angular
wave-number $\ell$. To this end, we compare the \emph{cosmic average}
power spectrum, 
\begin{equation}
  \label{eq:mapsignal}
  \mat{P}_{\rm s}(\ell):=
  \wtilde{\mat{W}}_{\ell}^{}\wtilde{\mat{S}}_{\ell}
   \wtilde{\mat{W}}_{\ell}^\dagger\;,
\end{equation}
of the reconstructed matter and galaxy-number density modes on the
lens planes to noise in the reconstruction from shape noise and tracer
sampling noise, which is
\begin{equation}
  \label{eq:mapnoise1}
  \mat{P}_{\rm n}(\ell):=
  \wtilde{\mat{W}}_{\ell}^{}
  \wtilde{\mat{X}}_{\ell}
  \wtilde{\mat{W}}_{\ell}^\dagger\;.
\end{equation}
Here we use the definitions
\begin{equation}
  \label{eq:fpfilter2}
  \wtilde{\mat{X}}_{\ell}:=
  \left(\wtilde{\mat{R}}_{\ell}^\dagger
    \wtilde{\mat{N}}_{\alpha\beta}^{-1}
    \wtilde{\mat{R}}_{\ell}^{}\right)^{-1}~;~
  \wtilde{\mat{W}}_{\ell}:=
  \wtilde{\mat{S}}_{\ell}\left(
    \wtilde{\mat{S}}_{\ell}+
    \wtilde{\mat{X}}_{\ell}\right)^{-1}\;.
\end{equation}
In this reconstruction, the Wiener filter $\wtilde{\mat{W}}_\ell$ uses
the true signal power $\wtilde{\mat{S}}_\ell$ present in the data. As
pointed out earlier, this is not a necessity but is required for an
optimal minimum-variance filter. For \mbox{$\alpha=\beta=0$} (neither
smoothing nor mixing), the noise covariance is \mbox{$\mat{P}_{\rm
    n}(\ell)=\wtilde{\mat{X}}_{\ell}$}.

\subsection{Galaxy-stochasticity noise}
\label{sect:gsn}

\ch{The noise covariance $\mat{P}_{\rm n}(\ell)$ contains only a part
  of the statistical uncertainty in a combined reconstruction; namely,
  this is the noise originating from the unknown intrinsic source
  galaxy shapes and galaxy sampling noise.} In the presence of
stochasticity between matter and tracer density, however, there is a
random scatter in the sample-noise-free tracer density for a given
matter density field \ch{that gives rise to the additional noise
  covariance $\mat{P}_{\rm gsn}$},
\begin{equation}
  \label{eq:mapnoise2}
  \mat{P}_{\rm n,all}(\ell):=
  \mat{P}_{\rm n}(\ell)+\mat{P}_{\rm gsn}(\ell)
\end{equation}
\citep[called random biasing field
in][]{1999ApJ...520...24D}. Contrary to Poisson shot-noise, this
\ch{galaxy-stochasticity noise (GSN)} is also present, if the number
of tracer galaxies were infinite. \ch{This is a new feature compared
  to reconstruction techniques relying only on the 3D lensing signal.
  Possible realisations of a galaxy-number density field for a given
  matter density field on the lens planes depend on the details of the
  physics behind the galaxy bias. Consequently, a precise estimate of
  the GSN level can only be provided if the galaxy bias scheme is
  known.}

\ch{For a first-order estimate of GSN, we assume Gaussian fluctuations
  in the galaxy number and matter density on every lens plane. In this
  Gaussian approximation, the bivariate p.d.f. of modes of the matter
  density contrast, $\tilde{\delta}^{(i)}_{\rm m}(\vec{\ell}),$ and a
  galaxy number density, $\tilde{n}^{(i)}_{\rm g}(\vec{\ell})$, are
  fully determined by the variance $P^{(i)}_\delta(\ell)$ of
  $\tilde{\delta}^{(i)}_{\rm m}$, the variance $[\bar{n}_{\rm
    g}^{(i)}b^{(i)}(\ell)]^2P^{(i)}_\delta(\ell)$ of
  $\tilde{n}^{(i)}_{\rm g}$, and the cross-correlation coefficient
  $r^{(i)}(\ell)$ of both. From this the variance of a galaxy tracer
  mode about a fixed matter density mode follows:}
\begin{equation}
  \label{eq:Pgcn}
  P_{\rm gsn}^{(i)}(\ell)=
  \left[\bar{n}_{\rm g}^{(i)}b^{(i)}(\ell)\right]^2
  \left(1-[r^{(i)}(\ell)]^2\right)P^{(i)}_\delta(\ell)\;.
\end{equation}
\ch{See Appendix \ref{ap:mockdatagauss} for details. The random
  biasing field is an independent Gaussian realisation with power
  spectrum $P_{\rm gsn}^{(i)}$.} The essential parameter for this
random scatter is $r(\ell)$, which vanishes for \mbox{$|r(\ell)|=1$},
but reaches a maximum in amplitude for \mbox{$r(\ell)=0$}.  On the
other hand, a smaller $r(\ell)$ also results in a reduction of the
mixing of matter and tracer density modes by the minimum-variance
filter. On the extreme end for \mbox{$r(\ell)=0$}, the filter does not
make use of any tracer information at all for the matter density
map. The total noise power per angular mode in the mass map can hence
then be approximated by
\begin{equation}
  \label{eq:powergcn}
  \mat{P}_{\rm gsn}(\ell):=
  \wtilde{\mat{W}}_{\ell}^{}
  \wtilde{\mat{S}}_\ell^{\rm gsn}
  \wtilde{\mat{W}}_{\ell}^\dagger\;,
\end{equation}
where
\begin{equation}
  \wtilde{\mat{S}}^{\rm gsn}_\ell:=
 {\rm diag}\left\{
    \underbrace{0,\ldots,0,}_{\rm N_{\rm lp}~elements}
    P^{(1)}_{\rm gsn}(\ell),\ldots,
    P^{(N_{\rm lp})}_{\rm gsn}(\ell)\right\}\;.
\end{equation}

For each lens plane, we translate these $\ell$-dependent GSN levels to
the noise variance on the map pixel scale by virtue of the integral
(STH09)
\begin{equation}
  \label{eq:pixelvar}
  \Ave{[\delta^{(i)}_{\rm gsn}(\Theta_{\rm s})]^2}=
  \int_0^\infty\frac{\d\ell\ell}{2\pi}
  |\wtilde{F}(\ell\Theta_{\rm s})|^2 [\mat{P}_{\rm gsn}(\ell)]_{ii}\;,
\end{equation}  
and likewise to compute $\ave{[\delta^{(i)}_{\rm m}(\Theta_{\rm
    s})]^2}$ for the signal power $[\mat{P}_{\rm s}(\ell)]_{ii}$ in
Eq. \Ref{eq:mapsignal}. \ch{The resulting ratios of GSN and signal
  power are}
\begin{equation}
  f^{(i)}_{\rm gsn}=
  \sqrt{\frac{\ave{[\delta^{(i)}_{\rm gsn}(\Theta_{\rm
        s})]^2}}{\ave{[\delta^{(i)}_{\rm m}(\Theta_{\rm s})]^2}}}:=
\frac{\sigma^{(i)}_{\rm gsn}}{\sigma^{(i)}_{\rm s}}
\end{equation}
for pixels in our fiducial mass map. For this estimate \ch{of
  $f^{(i)}_{\rm gsn}$}, we take the cosmic average
$P^{(i)}_\delta(\ell)$. This certainly underestimates the \ch{matter
  fluctuations} in galaxy cluster regions. On the other hand, the GSN
scales, at least for Gaussian random fields, linearly with the
amplitude of the actual matter density fluctuations, as in
Eq. \Ref{eq:Pgcn}, such that the ratio $f^{(i)}_{\rm gsn}$ should be a
robust approximation for Gaussian fields with more matter clustering.

\subsection{Correction for galaxy-stochasticity noise}
\label{sect:gsncorrect}

\ch{In practise, we estimate the S/N of the synergy reconstructions by
  dividing pixel values $\delta_{\rm m,est}(\vec{\theta})$ in the map
  by the pixel variance in noise realisations of the map. We obtain
  the noise realisations by randomising the source ellipticities and
  the tracer positions in accordance with their completeness and
  redshift errors $\mat{G}$. However, the noise realisations do not
  include the GSN but only contributions of $\sigma_{\rm shot}$ from
  galaxy shape- and tracer sampling noise.  In this section, we
  propose a GSN correction factor that is applied to this S/N map. The
  correction factor is based on the foregoing $f^{(i)}_{\rm gsn}$ and
  the variance $\sigma_{\rm shot}$ in the noise realisations.}

For each lens plane of the map the pixel variance $\sigma_{\rm all}^2$
has three independent components,
\begin{equation}
  \label{eq:sigmaall}
  \sigma^2_{\rm all}=
  \sigma^2_{\rm s}+\sigma^2_{\rm gsn}+\sigma^2_{\rm shot}
  =\left(1+f_{\rm gsn}^2\right)\sigma^2_{\rm s}+\sigma^2_{\rm shot}\;,
\end{equation}
where $\sigma_{\rm s}$ is the variance in the matter density signal,
$\sigma_{\rm gsn}$ is the GSN variance, and $\sigma_{\rm shot}$ is the
source shape- and tracer shot-noise variance. On the right hand side,
we have substituted the GSN variance by the signal variance and
$f_{\rm gsn}$. A S/N map that accounts for both $\sigma_{\rm shot}$
and $\sigma_{\rm gsn}$ is
\begin{eqnarray}
  \lefteqn{\frac{\delta_{\rm m,est}(\vec{\theta})}{\sqrt{\sigma^2_{\rm shot}+\sigma^2_{\rm gsn}}}=}\\
  &&\nonumber
  \frac{\delta_{\rm m,est}(\vec{\theta})}{\sigma_{\rm shot}}
  \left(1+\frac{\sigma^2_{\rm gsn}}{\sigma^2_{\rm shot}}\right)^{-1/2}=
  \frac{\delta_{\rm m,est}(\vec{\theta})}{\sigma_{\rm shot}}
  \times
  \left(1+f_{\rm gsn}^2\frac{\sigma^2_{\rm s}}{\sigma^2_{\rm shot}}\right)^{-1/2}\;,
\end{eqnarray}
where $\delta_{\rm m,est}(\vec{\theta})/\sigma_{\rm shot}$ on the
right hand side is the S/N invoking shot-noise only, \ch{as produced
  by randomising the catalogues.} For the correction factor inside the
brackets, the signal variance $\sigma_{\rm s}$ can be estimated by
employing Eq. \Ref{eq:pixelvar} with an appropriate Wiener-filtered
signal power spectrum.  In addition, the shot-noise variance,
$\sigma_{\rm shot}$, is determined by Eq. \Ref{eq:pixelvar} with the
Wiener filter noise power spectrum $\mat{P}_{\rm n}(\ell)$ inside the
integral. For a signal variance \mbox{$\sigma_{\rm s}\ll\sigma_{\rm
    shot}$}, the correction factor is roughly unity, which is always
the case for a cosmic average matter density power spectrum and our
fiducial survey. As we are mainly targeting galaxy cluster regions
with lensing cartography, however, a fiducial value of $\sigma_{\rm
  s}$ with higher variance than a cosmic average is likely. To obtain
a more realistic fiducial value, we construct an alternative signal
power spectrum for $\sigma_{\rm s}$, assuming \ch{(i) Gaussian
  fluctuations}, (ii) randomly scattered haloes with an average number
density $\bar{n}_{\rm sis}$ and (iii) haloes with \ch{an average
  singular isothermal sphere (SIS)} matter density profile (STH09):
\begin{equation}
  \label{eq:sismodel}
  \tilde{\delta}^{(i)}_{\rm sis}(\ell)=
  \frac{8\pi^2}{3\Omega_{\rm m}[1+z(\bar{\chi}_i)]}
  \left(\frac{\sigma_{\rm v}}{c}\right)^2
  \frac{D_{\rm H}^2}{f_{\rm K}(\bar{\chi}_i)\Delta\chi_i}
  \frac{1}{\ell}\;
\end{equation}
in Fourier space and SIS velocity $\sigma_{\rm v}$.
Therefore, the matter power spectrum for the $i$th lens plane is
described by 
\begin{equation}
  \label{eq:gsnmodelpower}
  P_\delta^{(i)}(\ell)=
  |\tilde{\delta}^{(i)}_{\rm sis}(\ell)|^2\bar{n}_{\rm sis}\;,
\end{equation}
which we insert into Eq. \Ref{eq:mapsignal} and Eq. \Ref{eq:pixelvar}
to calculate the pixel signal-variance $\sigma_{\rm s}$ (Appendix
\ref{sect:sismodel}).

\subsection{Cluster signal-to-noise}

\ch{We now consider the significance with which a single mass peak at
  a given radial distance can be detected in a synergy
  reconstruction. For a fiducial mass peak, we adopt a SIS-like matter
  over-density $\tilde{\delta}_{\rm sis}^{(i)}$ that is fully
  contained inside the $i$th lens plane, as in
  Eq. \Ref{eq:sismodel}. The associated average number density of
  tracers is on the level of a Gaussian approximation $\tilde{n}_{\rm
    sis}^{(i)}(\ell)=\overline{n}_{\rm
    g}^{(i)}r^{(i)}(\ell)b^{(i)}(\ell)\tilde{\delta}^{(i)}_{\rm
    sis}(\ell)$, which is analogous to the rationale in
  Sect. \ref{sect:rpsf}, and vanishes for all other lens planes $j\ne
  i$. When we combine this peaked mass model
  $\tilde{\vec{\delta}}_{\rm sis}(\ell)$ and the tracer density model,
  $\tilde{\vec{n}}_{\rm sis}(\ell)$, in $[\tilde{\vec{\delta}}_{\rm
    sis}(\ell),\tilde{\vec{n}}_{\rm sis}(\ell)]$, we acquire the
  average map response $[\overline{\vec{\delta}}_{\rm sis}(\Theta_{\rm
    s}),\overline{\vec{n}}_{\rm sis}(\Theta_{\rm s})]$ in a smoothed
  map by Eq. \Ref{eq:radialpsf}, where $\Theta_{\rm s}$ is the
  transverse smoothing kernel size.  The vector
  $\overline{\vec{\delta}}_{\rm sis}(\Theta_{\rm s})$ exhibits the
  expected mass map response to the central pixel of a SIS peak in the
  map.}

\ch{This signal is compared to the expected noise level inside a
  pixel. Relevant contributions to noise are (i) sample and shot
  noise, $\sigma_{\rm shot}^{(i)}$, (ii) the GSN variance $\sigma_{\rm
    gsn}^{(i)}$, and (iii) interference $\sigma^{(i)}_{\rm cn}$ by
  intervening matter density fluctuations on lens planes that do not
  host the fiducial SIS peak. The sources of noise (i) and (ii) are
  detailed in the Sect. \ref{sect:mapsn} and \ref{sect:gsn}. For (ii),
  we additionally assume that the interfering matter density power on
  all lens planes \mbox{$j\ne i$}, which do not host the SIS, is given
  by the cosmic average $P_\delta^{(j)}(\ell)$ in Eq. \Ref{eq:Pgcn},
  whereas we have Eq. \Ref{eq:gsnmodelpower} as a GSN model for the
  $i$th plane . We determine the pixel variance $\sigma_{\rm
    cn}^{(i)}$ in (iii) by the signal covariance $\mat{P}_{\rm
    s}(\ell)$, as noted in Eq. \Ref{eq:mapsignal}, whose diagonals
  $[\mat{P}_{\rm s}(\ell)]_{ii}$ are inserted into
  Eq. \Ref{eq:pixelvar}.  Finally, the radial S/N profile of the SIS
  peak in the map is $\delta_{\rm sis}^{(i)}(\Theta_{\rm
    s})/\sqrt{[\sigma^{(i)}_{\rm s}]^2+[\sigma^{(i)}_{\rm
      gsn}]^2+[\sigma^{(i)}_{\rm cn}]^2}$. As a theoretical S/N of the
  detection, we pick the lens plane index $i$ at maximum S/N, which
  may not correspond to the true distance of the SIS peak due to the
  $z$-shift bias.}

\section{Survey parameters}
\label{sect:mock}

\ch{We consider an idealised survey with homogeneous noise and a
  $\mat{G}$ that is independent of the pixel position to discuss the
  impact of a joint reconstruction in the following sections. This
  section defines the fiducial cosmology and binning details of the
  idealised survey. Moreover, we generate mock data to which the
  reconstruction algorithm is applied. The mocks utilise a N-body
  simulation of the large-scale dark matter density field populated
  with semi-analytical galaxies.}

\subsection{Fiducial parameters of the idealised survey}
\label{sect:improvement}

As fiducial cosmology, we use a standard flat $\Lambda$CDM model with
the matter-density parameter \mbox{$\Omega_{\rm m}=0.27$}, where
baryons are \mbox{$\Omega_{\rm b}=0.046$} and a shape parameter of
$\Gamma=0.19$. The normalisation of the matter fluctuations within a
sphere of radius $8\,h^{-1}\rm Mpc$ at a redshift of zero is
$\sigma_8=0.8$. For the spectral index of the primordial matter power
spectrum, we use $n_{\rm s}=0.96$.  \ch{With these parameters, we
  construct a fiducial 3D matter power spectrum according to
  \citet{Smith03} which is then used to model the signal covariance
  $\wtilde{\mat{S}}_\ell$.}

For the fiducial survey, we split the source galaxy catalogue into
\mbox{$N_z=20$} equally sized redshift slices of width \mbox{$\Delta
  z=0.1$}, which span the redshift range of \mbox{$z=0\ldots2$}.  For
the sources, we neglect the effect of redshift errors greater than the
width of the redshift slices, such that the true p.d.f. $p_z^{(i)}(z)$
of sources of the $i$th slice is well-described within
\mbox{$z\in[z_i,z_{i+1}]$} by the p.d.f. of redshift estimates of the
full sample,
\begin{equation}
  p_z(z)\propto z^2
  \exp{\left(-\left(\frac{z}{z_0}\right)^\lambda\right)}\;,
\end{equation}
where $z_0=0.57$, $\lambda=1.5$, and $z_i=(i-1)\Delta z$.  We
represent the reconstruction volume by \mbox{$N_{\rm lp}=10$} lens
planes between \mbox{$z=0$} and \mbox{$z=2$} that are centred within
slices of moderate width $\Delta z_{\rm lp}=0.2$. The total number
density of sources on the sky is \mbox{$\bar{n}=30\,\rm arcmin^{-2}$}
with an intrinsic shape noise of \mbox{$\sigma_\epsilon=0.3$}.

To support the matter density reconstruction, we include fiducial
galaxy tracers with known galaxy bias. For simplicity, their p.d.f.
of redshift estimates is identical to $p_z(z)$. Contrary to the
sources, however, we now also emulate the effect of redshift errors by
adopting a root-mean-square (r.m.s.)  accuracy of
\mbox{$\sigma_z(z)=0.04(1+z)$} (Gaussian errors), which is built into
$\mat{G}$ in Eq. \Ref{eq:gdef}. The slicing scheme for the tracers is
equivalent to the scheme of the sources. From this, we compute the
average number density of tracers $\bar{n}_{\rm g}^{(i)}$ and
$\bar{\eta}_{\rm g}^{(i)}$ from Eq. \Ref{eq:eta} for each redshift
slice, and the observed redshift distributions $p_{\rm f}^{(i)}(z)$ by
piecewise convolving the p.d.f. $p_z(z)$ with a Gaussian kernel of the
r.m.s. $\sigma_z(z)$.  For low redshifts, we have
\mbox{$\bar{\eta}_{\rm g}^{(i)}\approx \bar{n}_{\rm g}^{(i)}$}, but we
find differences at higher redshifts where \mbox{$\Delta z_{\rm
    lp}\approx\sigma_z$}. For the fiducial survey, we reduce the total
number of tracers to 10 percent of the sources, \mbox{$\bar{n}_{\rm
    g}=3\,\rm arcmin^{-2}$}, since a reconstruction realistically
requires a specifically selected tracer population for an accurately
known bias. Here, the tracers are clustered as matter with
\mbox{$b(\ell)=1$} for all redshifts, but, more relevantly, we assume
a slight stochasticity on all scales, namely, \mbox{$r(\ell)=0.8$}. A
high correlation, $r\gtrsim0.5$, for various galaxy populations is
expected from theoretical models \citep[e.g.,][]{2001MNRAS.321..439G}
and observed for some cases
\citep{2002ApJ...577..604H,2007A&A...461..861S,2012arXiv1202.6491J}.

\subsection{N-body mock data}
\label{sect:nbodymock}

\begin{figure}
   \begin{center}
     \psfig{file=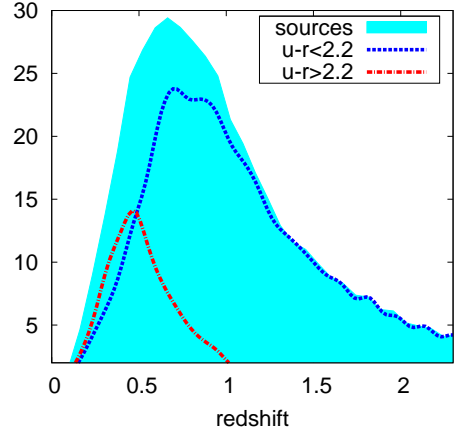,width=60mm,angle=-90}
   \end{center}
   \caption{\label{fig:nbodypofz} \ch{Distribution $\d N/\d z$ in
       units of galaxies per $\rm arcmin^2$ of three simulated galaxy
       samples in our N-body data. The data adopts a maximum depth of
       \mbox{$m_{\rm r}<25$} for all galaxies. The total galaxy sample
       used for the lensing analysis (sources) is further subdivided
       into red ($m_{\rm u}-m_{\rm r}>2.2$) and blue galaxies ($m_{\rm
         u}-m_{\rm r}\le2.2$).}}
\end{figure}

\begin{figure*}
   \begin{center}
     \psfig{file=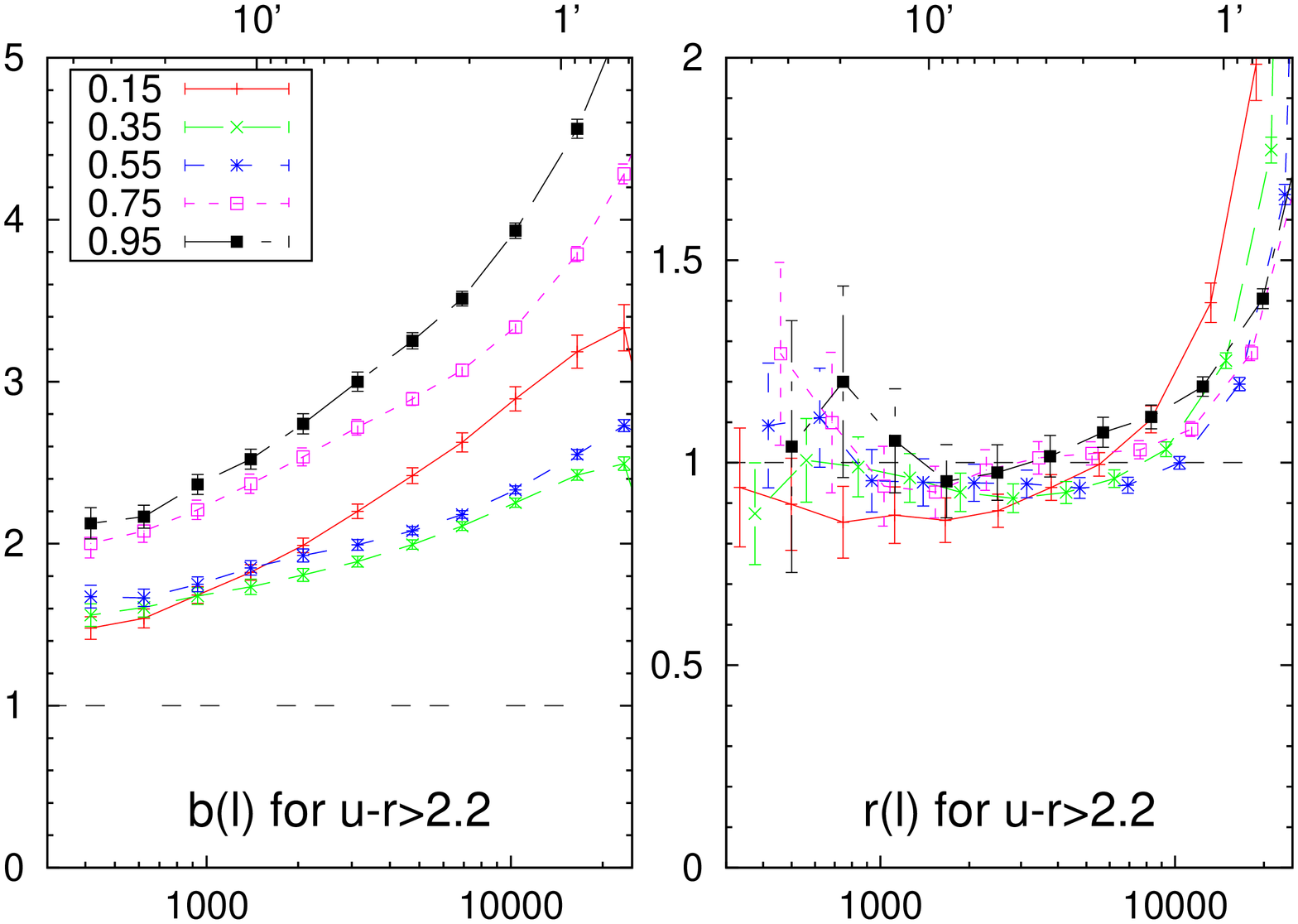,width=64mm,angle=-90}
     \psfig{file=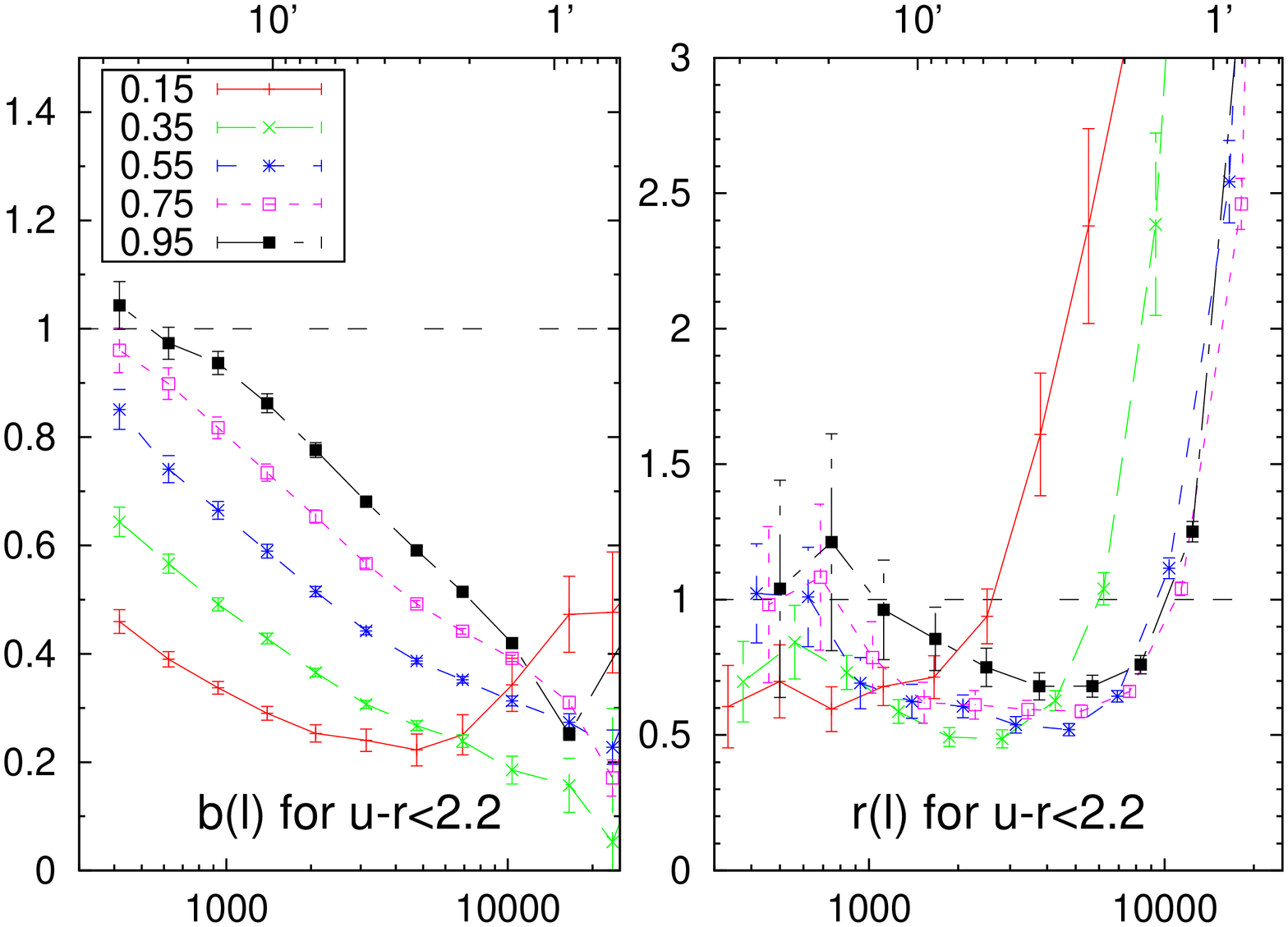,width=64mm,angle=-90}
   \end{center}
   \caption{\label{fig:nbodybias} \ch{Galaxy bias parameters $b(\ell)$
       and $r(\ell)$ in the Millennium Simulation data set as a
       function of angular scale $\ell$ ($x$-axis) and galaxy redshift
       (curves; see key for mean redshifts). The top numbers denote
       the corresponding aperture radius (arcmin) of the aperture
       statistics that were utilised to infer the galaxy bias from
       angular correlation functions (polynomial filter). The top two
       panels correspond to red galaxies with \mbox{$m_{\rm r}<25$}
       and \mbox{$m_{\rm u}-m_{\rm r}>2.2$} and the bottom two panels
       to blue galaxies with \mbox{$m_{\rm r}<25$} and \mbox{$m_{\rm
           u}-m_{\rm r}\le2.2$}. The error bars denote the remaining
       standard error based on 128 simulated survey fields with one
       square degree each.}}
\end{figure*}

\ch{For a realistic application of the methodology we employ the
  Millennium Simulation data set, a state-of-the-art N-body cold dark
  matter simulation with fiducial parameters of $\Omega_{\rm m}=0.25$,
  $\Omega_\Lambda=0.75$, $\Gamma=0.21$, $n_{\rm s}=1$, and
  $\sigma_8=0.9$ \citep{Millennium, 2005Natur.435..629S}. In the
  simulation, haloes of dark matter were populated with galaxies
  according to a semi-analytic recipe, as described in
  \citet{2011MNRAS.413..101G}\footnote{The galaxy properties have been
    obtained through the Millennium Simulation databases
    \citep{2006astro.ph..8019L}.}. We select galaxies with SDSS
  magnitudes of \mbox{$m_{\rm r}<25$} as a set of observable galaxies
  with known redshifts; Fig. \ref{fig:nbodypofz} displays the redshift
  distributions of all magnitude limited galaxy samples. For the
  simulated survey, we use galaxies from a $1\times1\,\rm\deg^2$ field
  and galaxies down to a redshift of $z=2$, yielding an average
  density of $\sim25$ sources per square arcmin. The mean redshift of
  the sources is $\bar{z}=1.0$.  Each source galaxy is equipped with a
  shear signal corresponding to its angular position and redshift. The
  shear signal is estimated by ray-tracing through a series of
  simulation snapshots in the direction of a source
  \citep{2009A&A...499...31H}. For the intrinsic shape noise we adopt
  a variance of the ellipticity of $\sigma_\epsilon=0.3$. We further
  subdivide the total galaxy sample into red ($m_{\rm u}-m_{\rm
    r}>2.2$) and blue galaxies ($m_{\rm u}-m_{\rm r}\le2.2$) to be
  used as galaxy tracers for the reconstruction technique of the mass
  map. We use only tracers below or equal $z=1$ to aid the
  reconstruction, which provides a density of $\sim10$ blue and
  $\sim5$ red tracers per square arcmin. For the mapping, all galaxy
  samples are split into redshift slices of width $\Delta z=0.1$
  within the regime \mbox{$0\le z<1$}, and a width of $\Delta z=0.2$
  within \mbox{$1\le z<2$} for the sources. Similar to the idealised
  fiducial survey we add Gaussian errors to the tracer redshifts with
  $\sigma_z(z)=0.04(1+z)$.}

\ch{The mapping methodology requires the specification of second-order
  galaxy bias parameters $\{b(\ell), r(\ell)\}$ of the tracer samples
  as a function of the angular scale $\ell$ and redshift.  We acquire
  estimators of the angular galaxy bias parameters by applying the
  methodology of \citet{1998ApJ...498...43S} and
  \citet{1998A&A...334....1V} to our simulated galaxy catalogues
  separately for each tracer redshift slice. We average the results
  thereof over all simulated 128 one-square-degree fields. Herein, we
  set the intrinsic shape noise to zero, as we do not attempt to
  account for uncertainties in bias parameters here. This lensing
  technique has already successfully been applied to real lensing
  data, as seen in \citet{2007A&A...461..861S}. We refer the reader to
  the latter article for the method details, which are irrelevant
  here. Figure \ref{fig:nbodybias} summarises the galaxy bias results
  of our tracer samples, including errorbars due to cosmic variance
  and sampling variance. In the following, we take the mean of all
  fields.  To determine these measurements, we employed, as in Simon
  et al. (2007), a polynomial filter for the aperture
  statistics. These statistics probe the angular second-order galaxy
  bias averaged over a $\ell$-band centred on $\ell_{\rm
    cen}\approx4.25/\theta_{\rm ap}$, where $\theta_{\rm ap}$ is the
  aperture radius in radians. The top $x$-axes values in the figure
  denote the values of $\theta_{\rm ap}$ that correspond to $\ell_{\rm
    cen}$ (bottom $x$-axes). The red tracers are more strongly
  clustered than matter, where \mbox{$b(\ell)>1$}, and highly
  correlated with the matter density field of $r(\ell)\approx 1$ on
  scales larger than a few arcmin. Blue tracers, on the other hand,
  are less clustered and less well correlated in both cases.}

\ch{The correlation factor $r(\ell)$ can exceed values of
  \mbox{$|r(\ell)|=1$}, because it is defined here and in the
  aforementioned references in terms of the tracer power spectrum
  $P_{\rm g}^{(i)}(\ell)$ from which the Poisson shot-noise
  $1/\bar{n}_{\rm g}^{(i)}$ has been subtracted. In the framework of a
  halo model and on scales dominated by haloes that are populated on
  average by \mbox{$\ave{N}<1$} galaxies, the shot-noise subtraction
  may lead naturally to \mbox{$r(\ell)>1$}, because galaxies can trace
  the matter distribution inside haloes by a sub-Poisson sampling
  process with a variance \mbox{$\ave{N(N-1)}^{1/2}<\ave{N}$}
  \citep{2001MNRAS.321..439G,2000MNRAS.318..203S}. The presence of
  central galaxies has a similar impact. We clearly observe this
  effect here for small angular scales in the simulation. The Wiener
  filters in Eq. \Ref{eq:combinedest} or Eq. \Ref{eq:combinedest2}
  diverge for \mbox{$|r(\ell)|>1$}, because the signal matrix
  $\mat{S}$ becomes singular. This indicates that our minimum-variance
  Ansatz, presuming sampling by a Poisson process, breaks down where
  the sub-Poisson effects become significant. To avoid this problem
  specific to small angular scales, we use more smoothing of
  $\Theta_{\rm s}=2\,\rm arcmin$ and clip correlation factors at
  $r(\ell)=0.9$. The latter affects the filter artificially by
  reducing the mixing for clipped modes and adding less information
  from the tracer clustering to the mass map. Note that we can always
  reduce the mixing inside the Wiener filter by adopting a lower
  correlation factor than in the data.}

\begin{figure*}
   \begin{center}
     \psfig{file=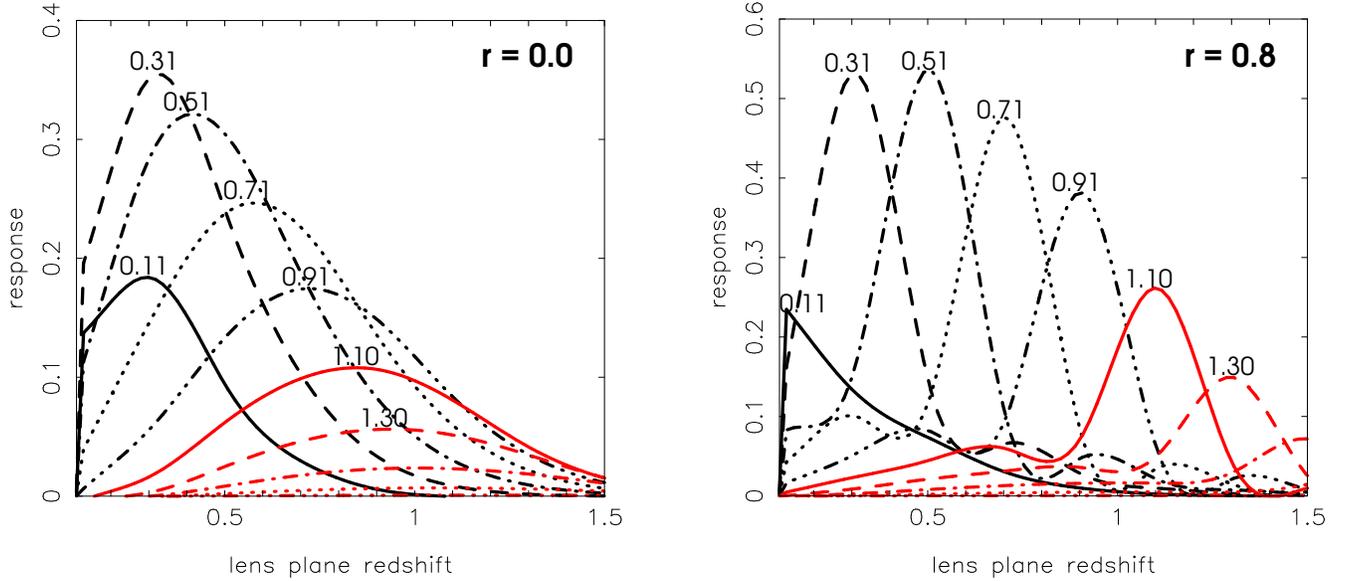,width=175mm,angle=0}
   \end{center}
   \caption{\label{fig:radialpsf} The radial p.s.f. in the smoothed
     mass maps for the case $r(\ell)=0.8$ (right panel) and no mixing
     of lensing and galaxy clustering ($r(\ell)=0$; left panel). The
     details of the fiducial survey are found in
     Sect. \ref{sect:improvement}. Both panels adopt $\alpha=0.01$ and
     $\beta=0.1$. The redshifts of the density peaks in the
     un-smoothed maps are the small number labels, which are only
     shown up to $z=1.3$.  The map pixel size is \mbox{$\Theta_{\rm
         s}=1\,\rm arcmin$}.}
\end{figure*}

\section{Results}
\label{sect:results}

\ch{In this section, we present our results for the S/N and radial
  p.s.f. in the idealised survey, and demonstrate the methodology for
  mock data based on a N-body simulation as blueprint for a realistic
  survey.}

\ch{For the idealised survey,} we set $\alpha=0.01$ to be consistent
with Fig. 11 of STH09 for a lensing-only reconstruction with a
transverse filter. Generally, the parameter $\alpha$ must not be too
close to unity, as this results in too much radial smoothing, which
moves basically all mass peaks to the middle of the reconstruction
volume (no radial information). \ch{Adjusting the tuning parameters
  below unity means that we scale the noise covariance towards less
  noise in the Wiener filter. Note that this does not mean that we
  obtain less noise in the reconstruction. In contrast, the Wiener
  filter applies less smoothing, which yields more noise in the map,
  but less bias.  For the synergy reconstruction, we adopt
  $\beta=0.1$.}  A parameter $\beta$ greater than $\alpha$ is a means
to down-weigh the impact of the tracers in the joint reconstruction,
\ch{which is desirable if the details of the galaxy bias are not
  accurately known.}

\subsection{Radial point spread function}

\begin{figure}
   \begin{center}
     \psfig{file=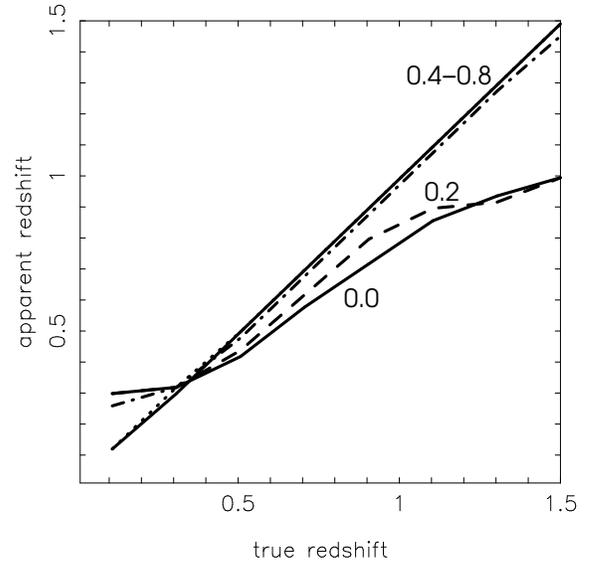,width=75mm,angle=0}
   \end{center}
   \caption{\label{fig:zshiftbias} Peak redshifts of the profiles of
     smeared mass peaks (ordinate) compared to the true peak redshifts
     (abscissa) for different correlation factors: \mbox{$r(\ell)=0$}
     (solid), \mbox{$r(\ell)=0.2$} (dashed), \mbox{$r(\ell)=0.4$}
     (dashed-dotted), and \mbox{$r(\ell)=0.6,0.8$} (indistinguishable
     diagonal lines).}
\end{figure}

The resulting p.s.f. of the \ch{idealised} fiducial survey and a pixel
size of \mbox{$\Theta_{\rm s}=1\,\rm arcmin$} is depicted in
Fig. \ref{fig:radialpsf} for the cases $r(\ell)=0,0.8$.  Owing to the
Wiener smoothing, the mass peaks are generally radially smeared, and
their amplitude is suppressed, especially for very small and high
redshifts. Compared to the lensing-only technique ($r(\ell)=0$),
however, adding tracers with \mbox{$r\ne0$} to the map-making process
clearly improves the p.s.f.: The radial profiles are narrowed and more
pronounced; the amplitudes are less suppressed. \ch{The peak maximum
  of the p.s.f. (apparent redshift) for given mass peak redshift (true
  redshift) determines the $z$-bias. The bias as a function of tracer
  correlation coefficient $r(\ell)$ is explored by
  Fig. \ref{fig:zshiftbias}. We essentially find no $z$-shift bias for
  \mbox{$r(\ell)\ge0.4$}.}

\begin{figure*}
   \begin{center}
     \epsfig{file=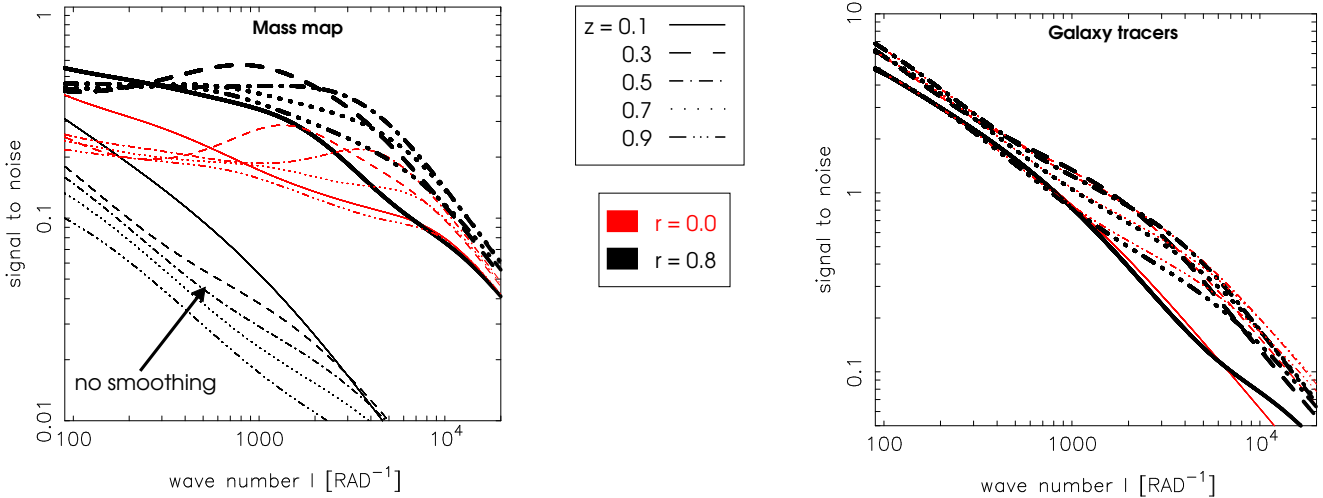,width=175mm,angle=0}
   \end{center}
   \caption{\label{fig:snpower} S/N of density modes for the matter
     density maps (left panel) and galaxy tracer density maps (right
     panel). The lines assume cosmic-average fluctuations in the
     density fields and $b(\ell)=1$ for the tracers. Different line
     styles correspond to different lens plane redshifts: $z=0.1$
     (solid), $z=0.3$ (dashed), $z=0.5$ (dashed-dotted), $z=0.7$
     (dotted), and $z=0.9$ (dashed-dotted-dotted).  The black thin
     lines do not employ any smoothing, while red thins lines use
     smoothing. Both use no mixing of lensing and galaxy clustering
     information. In the right panel, red and black thin lines
     coincide. Thick lines depict values in a Wiener smoothed map
     ($\alpha=0.01$, $\beta=0.1$) and a mixing with $r(\ell)=0.8$ for
     all redshifts. \ch{The S/N does not include GSN.}}
\end{figure*}

\subsection{Signal-to-noise of map}

\ch{For the idealised survey}, Fig. \ref{fig:snpower} depicts the S/N
of the lens plane density modes as a function of angular scale and
lens plane redshift. In the left panel, we have the matter density
modes; the right panel shows the tracer number density
modes. Different line styles correspond to different lens planes with
the thin lines to reconstructions with mode mixing switched off, or
$r(\ell)=0$, and the thick lines to the joint reconstruction. \ch{In
  addition, the black thin lines in the left panel depict the S/N in a
  map with no radial smoothing (\mbox{$\alpha=0$}) and no mixing.}
Clearly, a lensing-only map absolutely requires some radial smoothing,
which is seen here by comparing the low S/N of the thin black lines to
the boosted S/N in the red lines. The impact of a moderate mixing on
the S/N of the tracer number density maps (right panel) is small,
which is most prominently on the small angular scales. This changes
slightly if we choose an even larger tuning parameter $\beta$ (that is
not shown): A larger $\beta$ scales up the shot-noise of the tracers
inside the Wiener filter, which attributes even more weight to the
lensing data in the joint reconstruction. As the S/N of the tracers in
the data is actually higher than that of the shear, this will result
in a \emph{decreased} S/N for the galaxy-number density maps in
comparison to a reconstruction with no mixing; the joint
reconstruction is not optimal as to the map noise.

\subsection{Galaxy-stochasticity noise}

\begin{figure}
   \begin{center}
     \psfig{file=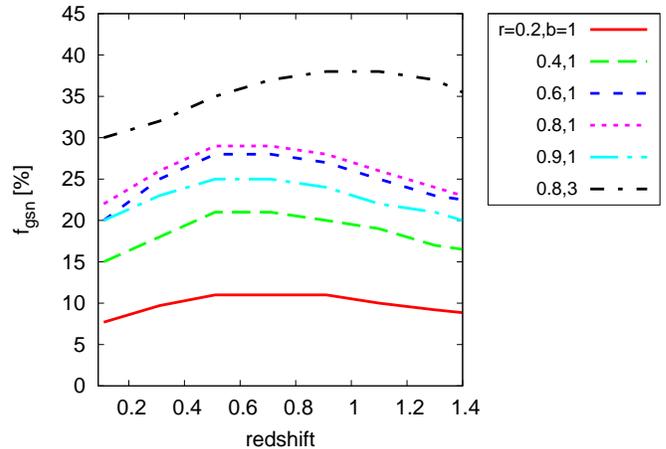,width=62mm,angle=-90}
   \end{center}
   \caption{\label{fig:gcnoise2} \ch{GSN variance $f_{\rm gsn}$ on a
       pixel scale of \mbox{$\Theta_{\rm s}=1\,\rm arcmin$} radius
       relative to the signal variance in the fiducial survey as
       function of lens plane redshift. The lines differ in their
       assumed $r(\ell)$. Except for one line, the bias factor of the
       tracers is $b(\ell)=1$. This figure uses $\alpha=0.01$ and
       $\beta=0.1$.}}
\end{figure}

\ch{Figure \ref{fig:gcnoise2} shows the estimated ratios $f_{\rm gsn}$
  of the pixel GSN-variance and pixel signal-variance for lens planes
  of increasing redshift. The map smoothing scale is $\Theta_{\rm
    s}=1\,\rm arcmin$.} We find the GSN \ch{on a pixel scale to be
  most prominent for \mbox{$r(\ell)\sim0.8$}}, which declines for
correlations greater or weaker than that; \ch{between
  $r(\ell)=0.6-0.8$, there is only little change, and in the absence
  of stochasticity, where \mbox{$r(\ell)=1$}, $f_{\rm gsn}$ vanishes.}
The dependence on lens plane redshift is marginal; most of the change
occurs below $z\lesssim0.4$. The GSN increases with the bias factor
$b(\ell)$ of the tracers. Overall, typical figures for $f^{(i)}_{\rm
  gsn}$ are below $30\%$, but can be above this level for strongly
clustered tracers.

\subsection{Cluster signal-to-noise}

\begin{figure}
   \begin{center}
     \psfig{file=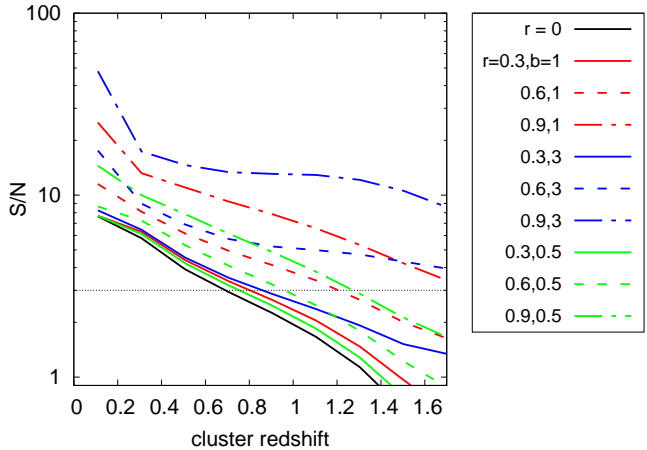,width=62mm,angle=-90}
   \end{center}
   \caption{\label{fig:clustersn} \ch{Trends in the signal-to-noise
       ratio for the detection of a SIS-like mass peak as function of
       the peak redshift ($\alpha=0.01$ and $\beta=0.1$); the pixel
       scale is \mbox{$\Theta_{\rm s}=1\,\rm arcmin$}. The mass of the
       peak corresponds to $M_{200}=6.6\times10^{14}\,M_\odot
       h^{-1}$. The S/N values scale roughly with $\propto
       M_{200}^{2/3}$. The different lines correspond to different
       fiducial values $\{r(\ell),b(\ell)\}$ of the bias of the galaxy
       tracers, as indicated inside the legend; \mbox{$r=0$} considers
       a lensing-only reconstruction. The horizontal black line
       indicates a $3\sigma$ detection.}}
\end{figure}

\ch{In Fig. \ref{fig:clustersn} we plot the S/N detection of a SIS
  mass peak as a function of peak redshift. The peak has the mass of a
  large galaxy cluster with
  $M_{200}=6.6\times10^{14}\,M_\odot\,h^{-1}$, or $\sigma_{\rm
    v}=10^3\,\rm km\,s^{-1}$, at $z=0$. Until redshift value
  $z\sim0.6$ this peak is visible above a $3\sigma$ limit when only 3D
  lensing information and a tuning of $\alpha=0.01$ are used; see the
  black solid line with \mbox{$r=0$}. The S/N scales with
  $M_{200}^{2/3}$ as discussed in STH09. The S/N detection improves
  when we combine the lensing information with the galaxy tracer
  information, adopting $\beta=0.1$; see lines with \mbox{$r>0$}.  The
  S/N improvement is greater for higher correlation factors $r(\ell)$
  or more clustering $b(\ell)$ of the tracers. Unless we have extreme
  cases of high correlations, where \mbox{$r\sim0.9$}, and strong
  clustering, where \mbox{$b\sim3$}, the S/N enhancement is only
  moderate between the factors of $2-3$. The GSN model adopts a
  Gaussian approximation with $\bar{n}_{\rm sis}=1\,\rm deg^{-2}$. For
  this approximation, shape noise and sampling noise are still the
  dominating source of pixel noise, such that a scaling of the S/N
  detection $\propto M_{200}^{2/3}$ is also found for the synergy
  technique within \mbox{$\sim10\%$} accuracy. We verified this within
  the mass range \mbox{$5\times10^{13}\,M_\odot h^{-1}\le M_{200}\le
    10^{15}\,M_\odot h^{-1}$}.}

\ch{When we consider both $r=0.8$ and $b=1$ for the GSN correction
  factor, we find that the S/N levels in a randomised map have to be
  reduced to $\sim75\%$ at $z=0.15$, $90\%$ at $z=0.25$, and
  $\gtrsim93\%$ at all other values
  (Sect. \ref{sect:gsncorrect}). These figures are typical values for
  $r(\ell)\in[0,1]$ and $b(\ell)\in[0,3]$. Therefore, the GSN is a
  small effect in the \emph{Gaussian regime} and mostly relevant at
  redshifts $z\lesssim0.3$.}

\subsection{N-body mock data}

\begin{figure*}
   \begin{center}
     \psfig{file=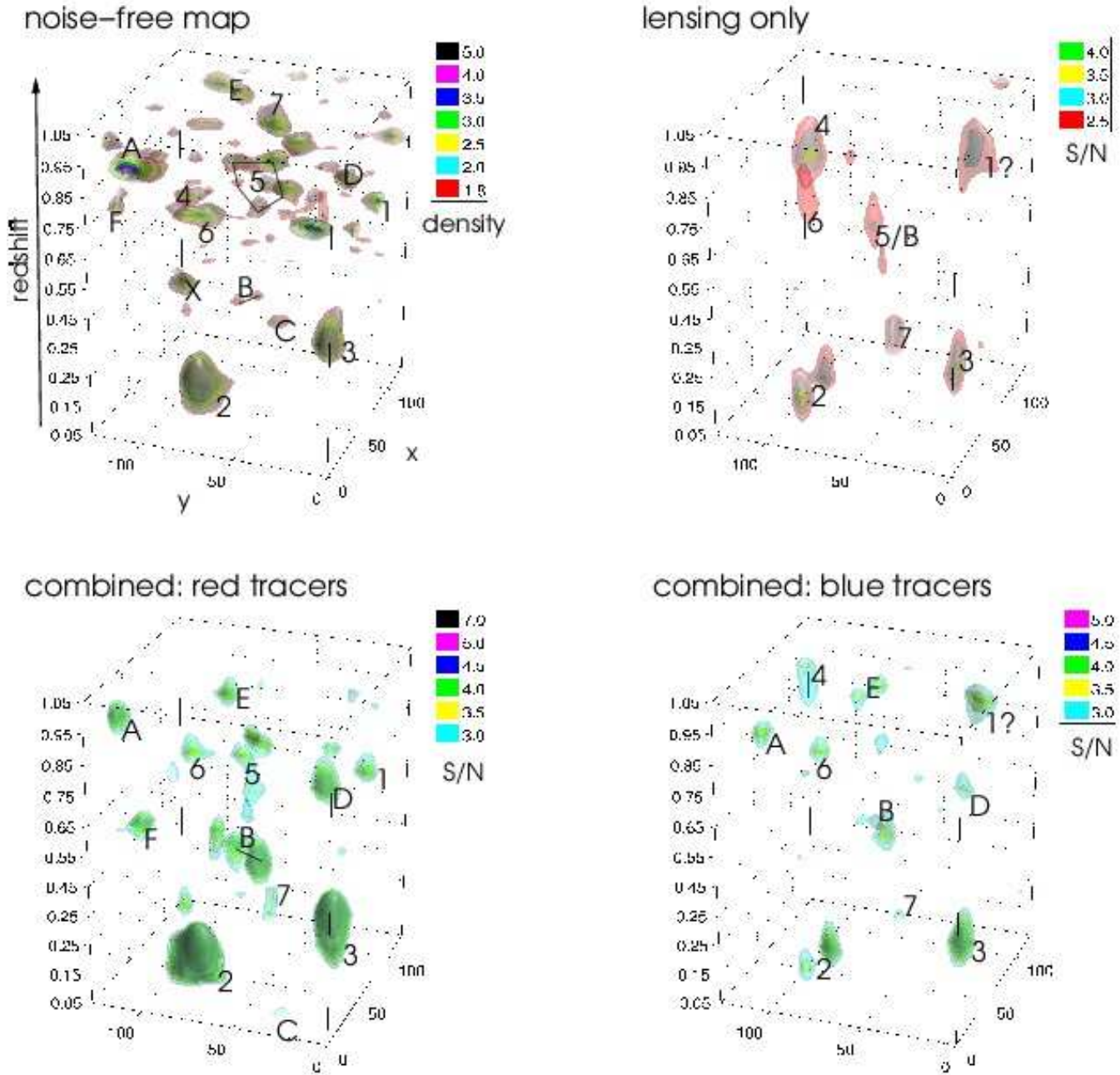,width=185mm,angle=0}
   \end{center}
   \caption{\label{fig:nbody} \ch{Simulated reconstructions of the
       original N-body data in the top left panel (density contrast
       iso-surfaces). The lens planes covering one square degree have
       sizes of $128\times128\,\rm pixel^2$ on the $x$- and
       $y$-axes. All maps are subject to smoothing (Gaussian kernel
       with two arcmin r.m.s. width). \emph{Top right}: Reconstruction
       using only the lensing information
       ($\alpha=0.05$). \emph{Bottom left}: Reconstruction adding
       information from red galaxy tracers with $m_{\rm u}-m_{\rm
         r}\ge2.2$ ($\alpha=0.05$, $\beta=0.5$). \emph{Bottom right}:
       Reconstruction including information from blue galaxies tracers
       with $m_{\rm u}-m_{\rm r}>2.2$ ($\alpha=0.05$,
       $\beta=0.5$). The number and capital letter labels indicate
       mass peak matches across different maps. The simulated
       reconstructions display signal-to-noise iso-surfaces. The
       combined maps do not account for GSN.}}
\end{figure*}

\ch{In Fig. \ref{fig:nbody}, we show the simulated mass mapping of one
  N-body simulated field in different versions. The field has been
  randomly selected from the set of 128 one square degree fields. The
  top left panel displays transparent matter density contrast
  iso-surfaces of the data without source galaxy shape noise and
  without synergy.  This ideal map has subsequently been transversely
  smoothed with a Gaussian kernel of 2 arcmin r.m.s. size. All maps in
  this figures are subject to the same angular smoothing. The top
  right panel corresponds to a map based on the lensing catalogue only
  but now with shape-noise of variance $\sigma_\epsilon=0.3$ and
  $\alpha=0.05$. This map and the following other two depict
  iso-surfaces of S/N based on 500 noise realisations that do not
  include GSN. The two bottom panels are the mass maps that include
  both the noisy lensing data and the information from galaxy tracers,
  which are either red galaxies with $m_{\rm u}-m_{\rm r}>2.2$ or blue
  galaxies with $m_{\rm u}-m_{\rm r}\le2.2$. The galaxy catalogues are
  flux-limited with $m_{\rm r}\le25.0$. In these maps, we set the
  tuning parameters to $\alpha=0.05$ and $\beta=0.5$. Mass peaks of
  more than $\sim3\sigma$ in the lensing-only map are designated by
  numbers between \emph{1} and \emph{6}. Mass peaks along the same
  line-of-sight and that are closest in redshift obtain the same
  number in the noise-free map. The distribution of mass peaks in the
  noise-free map was confusing and needed to be viewed on a computer
  display from different view angles to identify possible matches,
  especially at higher redshifts where a redshift slice of the light
  cone contains more volume. The complex \emph{5} comprises a series
  of peaks that are connected by lines to guide the eye. Significant
  mass peaks in the combined reconstruction that are not visible in
  the lensing-only map are given capital letters between \emph{A} and
  \emph{F}. Their possible matches are also indicated in the
  noise-free map. The question mark in \emph{1?}  indicates that the
  match to \emph{1} in the noise-free map is uncertain. By \emph{5/B},
  we mean that the peak is located between the peaks \emph{5} and
  \emph{B}, which are both along the same l.o.s. but at different
  redshifts. All maps recover the prominent structures \emph{2} and
  \emph{3} at low redshifts but fail to significantly recover \emph{C}
  and \emph{X}, which both appear prominent in the noise-free map. The
  benefit from adding tracer information is mostly visible at higher
  redshifts, at $z\gtrsim0.5$, where more individual structures are
  lifted above the $3\sigma$ threshold; peaks are less elongated in
  this regime as well. In particular, \emph{B} and \emph{5} are
  resolved when using red tracers, whereas the lensing-only map merges
  both together at an intermediate redshift. The blue tracers do not
  recover \emph{5} but at least \emph{B} at lower redshift. When using
  red tracers overall the S/N in the map is higher because red
  galaxies are more strongly clustered than blue galaxies and stronger
  correlated with the matter density field. The blue tracers render
  the original lensing-only map modestly in comparison by shifting the
  lensing signal in \emph{5/B} to \emph{B}, correcting \emph{6},
  weighing down \emph{7}, and adding a couple of new features of
  \emph{A}, \emph{D}, and \emph{E} that are insignificant in the
  lensing-only map. The redshift offset of peaks can be as high as
  \mbox{$\Delta z\approx0.2$}, as for seen for \emph{1?}  (blue), or
  \emph{F} (red).}

\section{Discussion}
\label{sect:discussion}

\ch{A synergy of 3D lensing data and galaxy clustering information can
  potentially alleviate the notorious $z$-shift bias in 3D lensing
  mass maps, provided the distribution of the tracers is statistically
  correlated with the underlying mass-density field.}  \ch{This can be
  seen in Fig. \ref{fig:radialpsf}, which compares the radial
  p.s.f. for uncorrelated tracers to the p.s.f. in a synergy
  reconstruction with highly correlated tracers. The synergy produces
  a mass map in which the p.s.f. now peaks on average at the redshift
  of the original mass peaks and in which mass peaks are less smeared
  out in radial direction (width of p.s.f). Moreover, the $z$-shift
  bias is already fixed for relatively loosely correlated tracers with
  \mbox{$r(\ell)>0.4$}, as the additional Fig. \ref{fig:zshiftbias}
  shows. Mixing the tracer and lensing information therefore promises
  to be an effective technique to address the $z$-shift bias.}

\ch{The 3D mass mapping with gravitational lensing is essentially a
  tool for the visualisation of the spatial distribution of mass peaks
  on a galaxy-cluster mass scale; a moderate synergy with tracers
  improves the accuracy of the distance estimates and the detection
  rate at greater distances.  Figure \ref{fig:clustersn} displays the
  change in S/N of cluster-sized mass peaks in a synergy map with
  moderate mixing (\mbox{$\alpha/\beta=0.1$}). In the case of
  \mbox{$r\gtrsim0.4$}, we expect a S/N enhancement by a factor
  $2-3$. Strongly clustered tracers with a bias of $b\sim3$ are an
  exception here as they yield even more enhancement. They, however,
  should not be utilised in a reconstruction, because large density
  fluctuations clearly cannot obey a Gaussian statistics, which is the
  underlying assumption of the GSN treatment in the figure; Gaussian
  density fluctuations, $\delta$, require a symmetric distribution
  about \mbox{$\delta=0$}, whereas large fluctuations
  \mbox{$\ave{\delta}^2\gg1$} are bound to have a skewed distribution
  due to the constraint \mbox{$\delta\ge-1$}. The S/N improvement at
  larger distances is underlined by Fig. \ref{fig:nbody}.  An increase
  in the S/N and less radial smearing, which is visible in
  Fig. \ref{fig:radialpsf}, at the same time results in a higher
  redshift accuracy of the mass peaks, because radial profiles of mass
  peaks are distinguishable more easily
  \citep{2012MNRAS.419..998S}. Therefore, the benefit from our new
  algorithm is also a higher redshift accuracy instead of a more
  complete visualisation of the spatial distribution of cluster-sized
  masses. Based on this, the search for lensing mass peaks can be
  supported by galaxy tracers, and lensing mass models of clusters can
  be refined by accounting for possible alignments of peaks close to a
  single l.o.s.}

\ch{Our synergy technique is linear and for this reason has limited
  applicability on sub-degree scales due to a potentially non-linear
  galaxy bias. The red thin lines in the left panel of
  Fig. \ref{fig:snpower} display the S/N of matter density modes in
  the 3D mass map before synergy. Compare this to the thin red lines
  in the right panel, which exhibit the S/N of the tracer number
  density modes that is roughly ten times higher. This basically
  quantifies the information on the matter density field as encoded in
  the tracer distribution, \emph{if} there is no stochastic galaxy
  bias and if the exact mapping between tracer and matter density is
  known \citep[deterministic galaxy bias; e.g.,
  ][]{1998MNRAS.293..209M}.  This seems to favour a large weight for
  the galaxy tracers in a synergy reconstruction, and this would
  result in a S/N boost compared to a lensing mass map. In reality,
  however, galaxy bias is stochastic, non-linear and possibly even
  non-local
  \citep[e.g.,][]{1999ApJ...518L..69T,2001ApJ...558..520Y,2002ApJ...577..604H,1999ApJ...520...24D,1999ApJ...525..543M},
  which is not properly accounted for in a linear filter: Our filter
  assumes by construction a linear relation between tracer and matter
  density, which is a Gaussian bias, and a Poisson process by which
  tracers sample the matter density field.  The former can be seen by
  the fact that only a linear mixing of both fields is possible within
  the filter. Gaussianity is a valid assumption on (smoothing) scales,
  where density fluctuations are small, \mbox{$\ave{\delta^2}\ll1$},
  or on large scales beyond \mbox{$\sim10\,\rm Mpc$}. Hence it is a
  fair assumption on angular scales larger than $\sim45$ arcmin (15
  arcmin) at \mbox{$z\sim0.2(0.8)$} but, on the other hand, is prone
  to bias the mass maps on smaller angular scales. Moreover,
  sub-Poisson sampling processes become relevant on small angular
  scales, as indicated by the shot-noise corrected correlation factor
  of \mbox{$r(\ell)>1$} in Fig. \ref{fig:nbodybias}. To reduce bias on
  these scales we weigh down the tracer information by adopting small
  values of $\alpha/\beta\lesssim0.1$ at the expense of map S/N, and
  we smooth the map with a kernel of several arcmin size.  To further
  relax this problem, it is also conceivable to exclude tracer
  information by setting \mbox{$r(\ell)=0$} at low redshifts of
  \mbox{$z\lesssim0.3$} where the lensing information is highest, as
  seen in Fig. \ref{fig:clustersn}. Despite these issues, we conclude
  that a moderate mixing and smoothing yields qualitatively sensible
  results from the reconstructions in Fig. \ref{fig:nbody} with two
  different tracer samples. Nevertheless, giving less weight to the
  tracers adds less information to the mass map, so that realistically
  only a modest S/N improvement is feasible with the synergy method on
  non-linear scales. In addition, there remains an uncertainty in the
  GSN due to non-linear stochastic galaxy bias that can only be
  quantified by more accurate modelling.}

\ch{In contrast to red galaxy tracers, blue galaxy tracers lead to
  modest but presumably more reliable improvements in 3D mapping.  We
  draw this conclusion from Fig. \ref{fig:nbody} that shows the
  combined reconstructions with red and blue galaxies in
  comparison. Clearly, including red galaxy clustering information
  adds more S/N to the map than blue galaxies. This can be explained
  by Fig. \ref{fig:clustersn} by considering that red galaxies are
  both more clustered and more strongly correlated with the matter
  density field in terms of $r(\ell)$ as seen
  Fig. \ref{fig:nbodybias}. On the other hand, the assumption of a
  Gaussian bias model is less appropriate for red galaxies than for
  blue galaxies because of their greater density fluctuations
  $\ave{\delta^2}$.  Typical blue galaxies exhibit density
  fluctuations smaller by a factor of $\sim5$ on arcmin scales, which
  in theory is even less than matter (\mbox{$b<1$}), and their number
  density is higher, which reduces the shot-noise error. Furthermore,
  blue galaxies are frequently field galaxies so that they also map
  out the large-scale matter distribution outside of clusters unlike
  red galaxies, which are preferentially found in galaxy clusters
  \citep{1984ApJ...281...95P,2011ApJ...736...59Z}. Considering the
  unknowns of the galaxy bias scheme, a blue tracer population is
  hence presumably the more favourable choice.}

\section*{Acknowledgements}

\ch{I thank Stefan Hilbert and Jan Hartlap for providing simulated
  galaxy and shear catalogues, based on the Millennium Simulation,
  which were utilised to demonstrate the reconstruction algorithm. The
  Millennium Simulation databases used in this paper and the web
  application providing online access to them were constructed as part
  of the activities of the German Astrophysical Virtual Observatory. I
  also acknowlegde the useful comments by Stefan Hilbert and the
  anonymous referee on the paper.}  The work in this was supported by
the European DUEL Research-Training Network (MRTN-CT-2006-036133) and
the Deutsche Forschungsgemeinschaft in the framework of the
Collaborative Research Center TR33 `The Dark Universe'.

\bibliographystyle{aa}
\bibliography{combine}

\begin{thebibliography}{44}
\expandafter\ifx\csname natexlab\endcsname\relax\def\natexlab#1{#1}\fi

\bibitem[{{Bacon} \& {Taylor}(2003)}]{bacontay}
{Bacon}, D.~J. \& {Taylor}, A.~N. 2003, MNRAS, 344, 1307

\bibitem[{{Bartelmann} \& {Schneider}(2001)}]{2001PhR...340..291B}
{Bartelmann}, M. \& {Schneider}, P. 2001, Physics Reports, 340, 291

\bibitem[{{Coles} \& {Jones}(1991)}]{1991MNRAS.248....1C}
{Coles}, P. \& {Jones}, B. 1991, \mnras, 248, 1

\bibitem[{{Dekel} \& {Lahav}(1999)}]{1999ApJ...520...24D}
{Dekel}, A. \& {Lahav}, O. 1999, \apj, 520, 24

\bibitem[{{Dodelson}(2003)}]{2003moco.book.....D}
{Dodelson}, S. 2003, {Modern cosmology}, ed. {Dodelson, S.}

\bibitem[{{Fan}(2003)}]{2003ApJ...594...33F}
{Fan}, Z. 2003, \apj, 594, 33

\bibitem[{{Guo} {et~al.}(2011){Guo}, {White}, {Boylan-Kolchin}, {De Lucia},
  {Kauffmann}, {Lemson}, {Li}, {Springel}, \& {Weinmann}}]{2011MNRAS.413..101G}
{Guo}, Q., {White}, S., {Boylan-Kolchin}, M., {et~al.} 2011, MNRAS, 413, 101

\bibitem[{{Guzik} \& {Seljak}(2001)}]{2001MNRAS.321..439G}
{Guzik}, J. \& {Seljak}, U. 2001, \mnras, 321, 439

\bibitem[{{Hilbert} {et~al.}(2009){Hilbert}, {Hartlap}, {White}, \&
  {Schneider}}]{2009A&A...499...31H}
{Hilbert}, S., {Hartlap}, J., {White}, S.~D.~M., \& {Schneider}, P. 2009, A\&A,
  499, 31

\bibitem[{{Hirata} \& {Seljak}(2004)}]{2004PhRvD..70f3526H}
{Hirata}, C.~M. \& {Seljak}, U. 2004, Phys. Rev. D, 70, 063526

\bibitem[{{Hoekstra} {et~al.}(2002){Hoekstra}, {van Waerbeke}, \&
  {Gladders}}]{2002ApJ...577..604H}
{Hoekstra}, H., {van Waerbeke}, L., \& {Gladders}, M.~D.~e. 2002, ApJ, 577, 604

\bibitem[{{Hu} \& {Keeton}(2002)}]{hukeeton02}
{Hu}, W. \& {Keeton}, C.~R. 2002, Phys. Rev. D, 66, 063506

\bibitem[{{Jullo} {et~al.}(2012){Jullo}, {Rhodes}, {Kiessling}, {Taylor},
  {Massey}, {Berge}, {Schimd}, {Kneib}, \& {Scoville}}]{2012arXiv1202.6491J}
{Jullo}, E., {Rhodes}, J., {Kiessling}, A., {et~al.} 2012,
  \texttt{arXiv:1202.6491}

\bibitem[{{Kaiser}(1992)}]{1992ApJ...388..272K}
{Kaiser}, N. 1992, \apj, 388, 272

\bibitem[{{Kaiser} \& {Squires}(1993)}]{1993ApJ...404..441K}
{Kaiser}, N. \& {Squires}, G. 1993, ApJ, 404, 441

\bibitem[{{Lemson} \& {Virgo Consortium}(2006)}]{2006astro.ph..8019L}
{Lemson}, G. \& {Virgo Consortium}, t. 2006, ArXiv Astrophysics e-prints

\bibitem[{{Leonard} {et~al.}(2012){Leonard}, {Dup{\'e}}, \&
  {Starck}}]{2011arXiv1111.6478L}
{Leonard}, A., {Dup{\'e}}, F.-X., \& {Starck}, J.-L. 2012, \aap, 539, A85

\bibitem[{{Mann} {et~al.}(1998){Mann}, {Peacock}, \&
  {Heavens}}]{1998MNRAS.293..209M}
{Mann}, R.~G., {Peacock}, J.~A., \& {Heavens}, A.~F. 1998, \mnras, 293, 209

\bibitem[{{Mart\'{i}nez} \& {Saar}(2002)}]{2002sgd..book.....M}
{Mart\'{i}nez}, V.~J. \& {Saar}, E. 2002, {Statistics of the Galaxy
  Distribution}, ed. {Mart\'{i}nez, V.~J.~\& Saar, E.} (Chapman and Hall/CRC)

\bibitem[{{Matsubara}(1999)}]{1999ApJ...525..543M}
{Matsubara}, T. 1999, \apj, 525, 543

\bibitem[{{Munshi} {et~al.}(2008){Munshi}, {Valageas}, {van Waerbeke}, \&
  {Heavens}}]{2008PhR...462...67M}
{Munshi}, D., {Valageas}, P., {van Waerbeke}, L., \& {Heavens}, A. 2008,
  \physrep, 462, 67

\bibitem[{{Pen} {et~al.}(2003){Pen}, {Lu}, {van Waerbeke}, \&
  {Mellier}}]{2003MNRAS.346..994}
{Pen}, U.-L., {Lu}, T., {van Waerbeke}, L., \& {Mellier}, Y. 2003, \mnras, 346,
  994

\bibitem[{{Postman} \& {Geller}(1984)}]{1984ApJ...281...95P}
{Postman}, M. \& {Geller}, M.~J. 1984, \apj, 281, 95

\bibitem[{{Schneider}(1998)}]{1998ApJ...498...43S}
{Schneider}, P. 1998, \apj, 498, 43

\bibitem[{{Schneider}(2006{\natexlab{a}})}]{2006glsw.conf....1S}
{Schneider}, P. 2006{\natexlab{a}}, in Saas-Fee Advanced Course 33:
  Gravitational Lensing: Strong, Weak and Micro, ed. {G.~Meylan, P.~Jetzer,
  P.~North, P.~Schneider, C.~S.~Kochanek, \& J.~Wambsganss}, 1--89

\bibitem[{{Schneider}(2006{\natexlab{b}})}]{2006glsw.conf..269S}
{Schneider}, P. 2006{\natexlab{b}}, in Saas-Fee Advanced Course 33:
  Gravitational Lensing: Strong, Weak and Micro, ed. {G.~Meylan, P.~Jetzer,
  P.~North, P.~Schneider, C.~S.~Kochanek, \& J.~Wambsganss}, 269--451

\bibitem[{{Seitz} \& {Schneider}(2001)}]{2001A&A...374..740S}
{Seitz}, S. \& {Schneider}, P. 2001, \aap, 374, 740

\bibitem[{{Seljak}(2000)}]{2000MNRAS.318..203S}
{Seljak}, U. 2000, \mnras, 318, 203

\bibitem[{{Simon}(2012)}]{2012arXiv1202.2046S}
{Simon}, P. 2012, \aap, 543, A2

\bibitem[{{Simon} {et~al.}(2007){Simon}, {Hetterscheidt}, {Schirmer}, {Erben},
  {Schneider}, {Wolf}, \& {Meisenheimer}}]{2007A&A...461..861S}
{Simon}, P., {Hetterscheidt}, M., {Schirmer}, M., {et~al.} 2007, A\&A, 461, 861

\bibitem[{{Simon} {et~al.}(2012){Simon}, {Heymans}, \&
  {Schrabback}}]{2012MNRAS.419..998S}
{Simon}, P., {Heymans}, C., \& {Schrabback}, e. 2012, \mnras, 419, 998

\bibitem[{{Simon} {et~al.}(2009){Simon}, {Taylor}, \&
  {Hartlap}}]{2009MNRAS.399...48S}
{Simon}, P., {Taylor}, A.~N., \& {Hartlap}, J. 2009, \mnras, 399, 48

\bibitem[{{Smith} {et~al.}(2003){Smith}, {Peacock}, {Jenkins}, \& {et
  al.}}]{Smith03}
{Smith}, R.~E., {Peacock}, J.~A., {Jenkins}, A., \& {et al.} 2003, MNRAS, 341,
  1311

\bibitem[{{Somerville} {et~al.}(2001){Somerville}, {Lemson}, {Sigad}, {Dekel},
  {Kauffmann}, \& {White}}]{2001MNRAS.320..289S}
{Somerville}, R.~S., {Lemson}, G., {Sigad}, Y., {et~al.} 2001, \mnras, 320, 289

\bibitem[{{Springel}(2005)}]{Millennium}
{Springel}, V. 2005, MNRAS, 364, 1105

\bibitem[{{Springel} {et~al.}(2005){Springel}, {White}, {Jenkins}, {Frenk},
  {Yoshida}, {Gao}, {Navarro}, {Thacker}, {Croton}, {Helly}, {Peacock}, {Cole},
  {Thomas}, {Couchman}, {Evrard}, {Colberg}, \& {Pearce}}]{2005Natur.435..629S}
{Springel}, V., {White}, S.~D.~M., {Jenkins}, A., {et~al.} 2005, Nat, 435, 629

\bibitem[{{Tegmark} \& {Bromley}(1999)}]{1999ApJ...518L..69T}
{Tegmark}, M. \& {Bromley}, B.~C. 1999, \apjl, 518, L69

\bibitem[{{Tegmark} \& {Peebles}(1998)}]{1998ApJ...500L..79T}
{Tegmark}, M. \& {Peebles}, P.~J.~E. 1998, \apjl, 500, 79

\bibitem[{{van Waerbeke}(1998)}]{1998A&A...334....1V}
{van Waerbeke}, L. 1998, \aap, 334, 1

\bibitem[{{VanderPlas} {et~al.}(2011){VanderPlas}, {Connolly}, {Jain}, \&
  {Jarvis}}]{2011ApJ...727..118V}
{VanderPlas}, J.~T., {Connolly}, A.~J., {Jain}, B., \& {Jarvis}, M. 2011, ApJ,
  727, 118

\bibitem[{{Weinberg} {et~al.}(2004){Weinberg}, {Dav{\'e}}, \&
  {Katz}}]{2004ApJ...601....1W}
{Weinberg}, D.~H., {Dav{\'e}}, R., \& {Katz}, N.~e. 2004, ApJ, 601, 1

\bibitem[{{Yoshikawa} {et~al.}(2001){Yoshikawa}, {Taruya}, \&
  {Jing}}]{2001ApJ...558..520Y}
{Yoshikawa}, K., {Taruya}, A., \& {Jing}, Y.~P.~e. 2001, ApJ, 558, 520

\bibitem[{{Zaroubi} {et~al.}(1995){Zaroubi}, {Hoffman}, \&
  {Fisher}}]{1995ApJ...449..446Z}
{Zaroubi}, S., {Hoffman}, Y., \& {Fisher}, K.~B.~e. 1995, ApJ, 449, 446

\bibitem[{{Zehavi} {et~al.}(2011){Zehavi}, {Zheng}, {Weinberg}, {Blanton},
  {Bahcall}, {Berlind}, {Brinkmann}, {Frieman}, {Gunn}, {Lupton}, {Nichol},
  {Percival}, {Schneider}, {Skibba}, {Strauss}, {Tegmark}, \&
  {York}}]{2011ApJ...736...59Z}
{Zehavi}, I., {Zheng}, Z., {Weinberg}, D.~H., {et~al.} 2011, \apj, 736, 59

\end{thebibliography}

\appendix

\section{Gaussian galaxy bias}
\label{ap:mockdatagauss}

\ch{Let $\tilde{\kappa}_{\rm g}=\tilde{n}_{\rm g}/\bar{n}_{\rm g}$ and
  $\tilde{\delta}_{\rm m}$ be the real part of the Fourier
  coefficients of the galaxy tracer number density contrast and matter
  density fluctuations, respectively, on a given lens plane and for a
  given angular mode $\vec{\ell}$. In the Gaussian regime the
  bivariate p.d.f. of both is given by}
\begin{equation}
  P\left(\tilde{\kappa}_{\rm g},\tilde{\delta}_{\rm m}\right)
  =\frac{1}{2\pi\sigma^2b\sqrt{1-r^2}}
  \exp{\left(-\frac{
        \tilde{\kappa}_{\rm g}^2/b^2+
        \tilde{\delta}_{\rm m}^2-
        2r\tilde{\kappa}_{\rm g}\tilde{\delta}_{\rm m}/b}
      {2\sigma^2(1-r^2)}\right)}\;,
\end{equation}
where the matter variance is $\sigma^2=\ave{\tilde{\delta}_{\rm
    m}^2}$, $\{b,r\}$ are the Gaussian bias parameters, and all means
$\ave{\tilde{\kappa}_{\rm g}}=\ave{\tilde{\delta}_{\rm m}}=0$
vanish. \ch{The same relation holds for the imaginary parts of the
  Fourier coefficients; the real and imaginary parts are independent.}
The conditional p.d.f.
\begin{equation}
  P\left(\tilde{\kappa}_{\rm g}\big|\tilde{\delta}_{\rm m}\right)=
  \frac{P\left(\tilde{\kappa}_{\rm g},\tilde{\delta}_{\rm m}\right)}
  {P\left(\tilde{\delta}_{\rm m}\right)}\;,
\end{equation}
is therefore also a Gaussian, namely with a mean of
\begin{equation}
\Ave{\tilde{\kappa}_{\rm g}\big|\tilde{\delta}_{\rm m}}=
\int\d\tilde{\kappa}_{\rm g}\,
P\left(\tilde{\kappa}_{\rm g}\big|\tilde{\delta}_{\rm m}\right)
\tilde{\kappa}_{\rm g}=
br\,\tilde{\delta}_{\rm m}
\end{equation}
and variance of
\begin{equation}
  \label{eq:galnoise}
  \sigma\left(\tilde{\kappa}_{\rm g}\big|
    \tilde{\delta}_{\rm m}\right)=
  \int\d\tilde{\kappa}_{\rm g}\,
  P\left(\tilde{\kappa}_{\rm g}\big|\tilde{\delta}_{\rm m}\right)
  \tilde{\kappa}^2_{\rm g}
  =b\sqrt{1-r^2}\,\sigma\;.
\end{equation}
The variance in $\tilde{\delta}_{\rm m}$ is \ch{given by the matter
  power spectrum $P_\delta(\ell)$ for the lens plane, the solid angle
  $A_{\rm fov}$ of the plane, and}
\begin{equation}
  \label{eq:powern}
  \sigma^2=
  \frac{P_\delta(\ell)}{2A_{\rm fov}}\;.
\end{equation}
\ch{Therefore, we expect an average tracer number density of
  $br\,\tilde{\delta}_{\rm m}$ with r.m.s. variance
  $b\sqrt{1-r^2}\sigma$ for a fixed matter density mode
  $\tilde{\delta}_{\rm m}$. The latter gives rise to the GSN in the
  Gaussian case \citep{1999ApJ...520...24D}.}

\section{SIS power spectrum}
\label{sect:sismodel}

A set of $N_{\rm h}$ haloes with positions $\vec{\theta}_j$ on the
$i$th lens plane and an average, axial-symmetric matter density
contrast, $\delta_{\rm sis}(|\vec{\theta}|)$, produces the combined
density contrast
\begin{equation}
  \delta_{\rm m}^{(i)}(\vec{\theta})=
  \sum_{j=1}^{N_{\rm h}}
  \delta_{\rm sis}(|\vec{\theta}-\vec{\theta}_j|)~;~
  \tilde{\delta}_{\rm m}^{(i)}(\vec{\ell})=
  \tilde{\delta}_{\rm sis}(\ell)\,
  \sum_{j=1}^{N_{\rm h}}{\rm e}^{+{\rm i}\vec{\ell}\vec{\theta}_j}\;,
\end{equation}
where the second equation on the right hand side is the Fourier
transform of $\delta_{\rm m}^{(i)}(\vec{\theta})$. Averaging the
two-point correlator of the density in Fourier space over all halo
positions results in
\begin{eqnarray}
 \lefteqn{\Ave{\tilde{\delta}_{\rm
       m}^{(i)}(\vec{\ell}_1)\tilde{\delta}_{\rm m}^{(i)}(\vec{\ell}_2)}=}\\
 &&\nonumber
  \tilde{\delta}_{\rm sis}(\ell_1)\tilde{\delta}_{\rm
    sis}(\ell_2)
  \left[\sum_{j=1}^{N_{\rm h}}\Ave{{\rm e}^{+{\rm
        i}(\vec{\ell}_1+\vec{\ell}_2)\vec{\theta}_j}}
  +
  \sum_{j\ne k=1}^{N_{\rm h}}    
  \Ave{{\rm e}^{+{\rm
        i}\vec{\ell}_1\vec{\theta}_j}
    {\rm e}^{+{\rm
        i}\vec{\ell}_2\vec{\theta}_k}}
  \right]\;.
\end{eqnarray}
We ignore the clustering of the haloes over the field-of-view $A_{\rm
  fov}$, so that the two-halo term in the second sum vanishes, and
\begin{eqnarray}
  \Ave{\tilde{\delta}_{\rm m}^{(i)}(\vec{\ell}_1)\tilde{\delta}_{\rm m}^{(i)}(\vec{\ell}_2)}&=&
  \tilde{\delta}_{\rm sis}(\ell_1)\tilde{\delta}_{\rm
    sis}(\ell_2)
  \sum_{j=1}^{N_{\rm h}}\frac{(2\pi)^2}{A_{\rm fov}}
  \delta_{\rm D}(\vec{\ell}_1+\vec{\ell}_2)\\
  &=&
  (2\pi)^2
  \delta_{\rm D}(\vec{\ell}_1+\vec{\ell}_2)
  \,\tilde{\delta}_{\rm sis}(\ell_1)\tilde{\delta}_{\rm
    sis}(\ell_2)
  \bar{n}_{\rm sis}\\
  &=&(2\pi)^2\delta_{\rm D}(\vec{\ell}_1+\vec{\ell}_2)
  \,P_\delta^{(i)}(\ell_1)\;,
\end{eqnarray}
where $\bar{n}_{\rm sis}:=N_{\rm h}/A_{\rm fov}$ expresses the mean
number density of haloes. Therefore, we obtain in this scenario
\begin{equation}
  P^{(i)}_\delta(\ell)=
  |\tilde{\delta}_{\rm sis}(\ell)|^2\bar{n}_{\rm sis}\;.
\end{equation}

\end{document}